# THE RENORMALIZATION OF COMPOSITE OPERATORS IN YANG-MILLS THEORIES USING GENERAL COVARIANT GAUGE *

John C. Collins †

and

Randall J. Scalise ‡

Department of Physics
104 Davey Laboratory
The Pennsylvania State University
University Park, PA 16802

**Abstract**

Essential to QCD applications of the operator product expansion, etc., is a knowledge of those operators that mix with gauge-invariant operators. A standard theorem asserts that the renormalization matrix is triangular: Gauge-invariant operators have 'alien' gauge-variant operators among their counterterms, but, with a suitably chosen basis, the necessary alien operators have only themselves as counterterms. Moreover, the alien operators are supposed to vanish in physical matrix elements. A recent calculation by Hamberg and van Neerven apparently contradicts these results. By explicit calculations with the energy-momentum tensor, we show that the problems arise because of subtle infra-red singularities that appear when gluonic matrix elements are taken on-shell at zero momentum transfer.

*Available by anonymous ftp at ftp.phys.psu.edu:pub/preprint/psuth141.ps.Z.uu

†e-mail: collins@phys.psu.edu
‡e-mail: scalise@phys.psu.edu

## 1 Historical Introduction

Much phenomenology in QCD requires the use of the operator product expansion [1–3] and many generalizations such as 'factorization theorems' [4]. Among the ingredients are matrix elements of particular gauge-invariant operators, which correspond to parton densities (or distribution functions). The properties of these operators under renormalization are vital to all QCD calculations, and one serious complication arises because gauge-invariant operators mix[1] with certain gauge-variant (non-gauge-invariant) operators. The renormalization directly determines the phenomenologically important anomalous dimensions of the operators—generally used in the form of Altarelli-Parisi splitting coefficients.

The extra operators that mix with the gauge-invariant operators are unphysical—we will call them 'alien' operators. It has been known since the earliest days of QCD that one must demonstrate that these alien operators do not contribute to physics. Three theorems apply to the decoupling: One is that a basis can be chosen such that the alien operators are BRST-exact. Next, physical matrix elements of BRST-exact operators are zero. The last theorem is a trivial consequence of the second: The renormalization mixing matrix is triangular—alien operators do not mix with the physical operators. The theorems to establish this have been proven in their strongest form by Joglekar and Lee [5] and more recently by Henneaux [6].

Unfortunately, recent calculations by Hamberg and van Neerven [7, 8] contradict these general theorems. Their results therefore throw into doubt the basis of all higher order perturbative QCD calculations. Our purpose in this paper is to resolve this contradiction between the theorem and the calculations. We will show that the contradiction is only apparent, and that it arises from certain subtle infra-red (IR) problems that are unluckily intrinsic to the usual algorithms for doing perturbative QCD calculations. However, the problem of efficiently performing practical calculations is left for future work.

The immediate motivation for the calculations by Hamberg and van Neerven was a long-standing discrepancy between calculations of the two-loop anomalous dimensions of the twist-2 covariant gluon operators in Feynman gauge [9–11] and the light-like axial gauge [12]. Since these anomalous dimensions are measurable, calculations performed in different gauges should agree, and this can readily be shown by the methods of [13], provided that one assumes the Joglekar–Lee theorem.

[1] Multiplicative renormalization is not sufficient to remove the infinities from Green functions of arbitrary composite operators; counterterms corresponding to different operators are needed.

Hamberg and van Neerven repeated the Feynman gauge calculation and discover that the older calculations [9–11] are in error because they assumed the applicability of the theorem that the renormalization matrix is triangular. Hamberg and van Neerven show that the renormalization matrix appears to be non-triangular. Their calculation supports the otherwise dubious light-cone gauge result and is in accord with supersymmetry arguments [8].

The roots of this failure are already present in the one-loop part of the calculation. Although Hamberg and van Neerven do not remark on it, their calculation shows that the finite part of a physical matrix element of their alien operator is nonzero at one-loop order, contradicting the second of the Joglekar–Lee results mentioned previously. They perform their calculation for the whole tower of twist-2 covariant gluon operators, but the problems are present for the simplest of these operators, the energy-momentum tensor $\theta_{\mu\nu}$, for which the renormalization theory was worked out by Freedman, Muzinich and Weinberg [14, 15]. The alien operators in that analysis are not manifestly the same as those used by Hamberg and van Neerven. The form of $\theta_{\mu\nu}$ given by Freedman *et al.* is in agreement with the general theorems of Joglekar and Lee and of Henneaux. However, the gauge-variant operators used by Hamberg and van Neerven are obtained from the analysis of Dixon and Taylor; it is not evident that these operators are BRST-exact.

This is where we start: A sufficiently detailed analysis of the energy-momentum tensor at one-loop order is enough to locate the source of the contradiction. We will verify that at one-loop order, the renormalization given by Freedman *et al.* is in fact correct. However, the one-loop gluonic matrix element of the alien operator fails to vanish at zero momentum transfer.[2, 3] We will find that the source of this incongruity is an infra-red divergence, but the divergence is not in the calculation of the matrix element. Rather, it is a quadratic divergence in the *proof* that the matrix element vanishes.

[2]The momentum transfer is defined to be the sum of the momenta flowing into the inserted operator vertex.

[3]Harris and Smith in a recent preprint [16] calculate a nonzero gauge *dependence* for the gluonic matrix element of the gauge-invariant part of $\theta_{\mu\nu}$ at *nonzero* momentum transfer. This is again in contradiction with the general theorems.

The source of the divergence makes it clear that the proof of the theorem on the vanishing of the alien operators should be correct when one applies it to physical states. The problem arises when one considers matrix elements in an off-shell gluon state and then takes the gluon on-shell.

But this clearly threatens the rationale for the usual methods of doing perturbative QCD calculations. Moreover, the renormalization matrix that Hamberg and van Neerven calculated and found to be non-triangular presumably includes some infra-red renormalization, contrary to what should be done.

In Section 2, we state the Joglekar–Lee theorems. In Section 3, we list our conventions for pure-gauge Yang-Mills theory. In Section 4, we give the results of the one-loop calculation of two-point Green functions with the energy-momentum tensor operator inserted at zero momentum transfer and derive the renormalization mixing matrix. The calculation at nonzero momentum transfer is currently underway. The off-shell results, as well as the physical matrix elements, will be reported in the near future. The Appendix contains: a brief discussion of 'right derivatives'; the full Lorentz tensor structure of the two-gluon Green functions, which are abbreviated in the text, with a separation into leading- and higher-twist pieces; a list of the Feynman graphs with composite operator insertions used in the calculations; and the Feynman rules for the operator vertices considered in this paper.

## 2 Renormalization of Gauge-Invariant Operators

In this section, we state the three theorems that Joglekar and Lee [5] proved on the renormalization of gauge-invariant operators. In [13], the theorems are stated and all the easy parts are proven.

Let $G_i$ denote a set of gauge-invariant operators that mix under renormalization, and let $A_i$ denote the set of alien operators with which they mix under renormalization. (We define 'alien' to mean 'not gauge-invariant'.) Finally, let $E_i$ denote the set of operators that vanish by use of the equations of motion[4] and with which the previous two sets of operators mix under renormalization.

The first of the Joglekar–Lee results is that the basis of the alien operators $A_i$ that mix with gauge-invariant operators can be chosen so that they are all BRST-exact, i.e., they can be written as[5]

$$A_i \approx \delta_{\rm BRST} B_i \quad , \tag{2.1}$$

where we will call $B_i$ the 'ancestor' of $A_i$.

The second theorem is that physical matrix elements of the BRST-exact alien operators, $\delta_{\rm BRST} B_i$, are zero.

The last of the theorems is that the renormalization mixing matrix is triangular:

$$\begin{pmatrix} R[G] \\ R[A] \\ R[E] \end{pmatrix} = \begin{pmatrix} Z_{GG} & Z_{GA} & Z_{GE} \\ 0 & Z_{AA} & Z_{AE} \\ 0 & 0 & Z_{EE} \end{pmatrix} \begin{pmatrix} G \\ A \\ E \end{pmatrix} . \tag{2.2}$$

Of these theorems, the hardest to prove is the first. It can easily be shown that all counterterms to BRST-invariant operators are themselves BRST-invariant [13]. Then one must prove that any BRST-invariant operator is a linear combination of gauge-invariant operators and BRST-exact operators. Up to operators that vanish by the equations of motion, this is supposed to be proven by Joglekar and Lee [5], but we find that proof very hard to understand. A simpler proof on the basis of cohomology theory is presented by Henneaux in [6].

A simple proof of the last two theorems is given in [13]. The vanishing of physical matrix elements of the alien operators follows from a simple Ward identity involving the BRST variation

[4] Matrix elements of $E_i$ must vanish, but Green functions of $E_i$ do not.

[5] We use the convention of [17], where the wavy equal sign means that the relation is only true after one or more of the equations of motions have been used.

(also called a Slavnov-Taylor identity), once one knows that only BRST-exact operators are needed. This result trivially generalizes to show that Green functions of these alien operators with BRST-invariant operators are zero. BRST-invariant operators include gauge-invariant operators and the BRST-exact operators that comprise all our alien operators.

The third theorem, on the triangularity of the renormalization matrix, immediately follows [13]. If any on-shell, physical matrix element of an *unrenormalized* operator in class $A$ is to vanish, then its pole piece must also vanish on shell.[6] Since at least some of the physical matrix elements of any gauge-invariant operator are non-vanishing, it follows that the entry $Z_{AG}$ must be zero; no operators in class $G$ can mix with the operator in class $A$. Similarly, $Z_{EG}$ and $Z_{EA}$ in the bottom row of the mixing matrix must be zero because an *unrenormalized* operator in class $E$ must vanish by the equations of motion, therefore its pole part must also vanish by the equations of motion.

Note that at the level of pure Feynman graph calculations, a physical matrix element is one with the gluon polarizations being purely transverse and with the states being on-shell quarks or on-shell gluons.

Prior to the work of Joglekar and Lee, it was shown by Freedman, Muzinich, and Weinberg [14, 15] how to construct a finite energy-momentum tensor for gauge theories. Their operator can readily be seen to satisfy the first Joglekar–Lee theorem, as we will explain later.

The problem we now face is that the calculations by Hamberg and van Neerven appear to violate all of the above theorems.

## 3 Yang-Mills Conventions

In this section, we list some common objects in Yang-Mills theory to exhibit our conventions and notation, but also because some, such as the energy-momentum tensor, play a pivotal rôle throughout this article.

[6]'Pole' in this context means a singularity as the dimension of space-time is varied. We are assuming the use of minimal subtraction with dimensional regularization to perform the renormalization (see Section 3.5).

## 3.1 Lagrangian Density

The effective Lagrangian density of pure-gauge Yang-Mills theory in general covariant gauge is, in terms of unrenormalized (bare) fields and parameters (designated by hats),

$$\mathcal{L}(x) = -\frac{1}{4}\hat{F}_a^{\mu\nu}(x)\hat{F}_{\mu\nu\; a}(x) - \frac{1}{2}\hat{\lambda}[\partial\cdot\hat{A}_a(x)][\partial\cdot\hat{A}_a(x)] + [\partial^\mu\hat{\eta}_a(x)][\hat{D}_\mu(x)\hat{\omega}(x)]_a, \qquad (3.1)$$

where the antisymmetric field strength tensor is given by

$$\hat{F}_a^{\mu\nu}(x) \equiv \partial^\mu\hat{A}_a^\nu(x) - \partial^\nu\hat{A}_a^\mu(x) - \hat{g}c_{abc}\hat{A}_b^\mu(x)\hat{A}_c^\nu(x) \qquad (3.2)$$

and the covariant derivative acts on fields in the adjoint representation of the group as follows

$$[\hat{D}_\mu(x)\hat{\omega}(x)]_a \equiv \hat{D}_{\mu\; ac}(x)\hat{\omega}_c(x) \equiv [\partial_\mu\delta_{ac} - \hat{g}c_{abc}\hat{A}_{\mu\; b}(x)]\hat{\omega}_c(x) = \partial_\mu\hat{\omega}_a(x) - \hat{g}c_{abc}\hat{A}_{\mu\; b}(x)\hat{\omega}_c(x). \qquad (3.3)$$

We are defining the Grassmann field $\hat{\eta}_a(x)$ to be the anti-ghost and the Grassmann field $\hat{\omega}_a(x)$ to be the ghost. The parameter $\hat{g}$ is the coupling strength, $c_{abc}$ are the structure constants of the underlying Lie algebra $SU(N)$, and $\hat{\lambda}$ is the arbitrary gauge-fixing parameter in general covariant gauge. The color indices in the adjoint representation $a, b, c, \ldots$ range from 1 to $N^2 - 1$.

## 3.2 Euler-Lagrange Equations of Motion

The Euler-Lagrange equations of motion, using right derivatives for the Grassmann variables[7] are

$$\frac{\partial^r\mathcal{L}_{\text{eff}}}{\partial\Phi_a} - \partial_\mu\frac{\partial^r\mathcal{L}_{\text{eff}}}{\partial(\partial_\mu\Phi_a)} = 0, \qquad (3.4)$$

where

$$\Phi_a \;\in\; \{\hat{A}_{\nu\; a}, \hat{\omega}_a, \hat{\eta}_a\}. \qquad (3.5)$$

[7]See Section A.1 in the Appendix for a discussion of right derivatives.

We have

$$(\hat{D}_\mu \hat{F}^{\mu\nu})_a + \hat{\lambda}\partial^\nu \partial \cdot \hat{A}_a + \hat{g} c_{abc}(\partial^\nu \hat{\eta}_b)\hat{\omega}_c = 0, \tag{3.6a}$$

$$(\hat{D}_\mu \partial^\mu \hat{\eta})_a = 0, \tag{3.6b}$$

$$\partial^\mu (\hat{D}_\mu \hat{\omega})_a = 0. \tag{3.6c}$$

### 3.3 BRST Symmetry

The gauge-fixed effective Lagrangian density is not gauge-invariant but is invariant under the following global symmetry [18]

$$\begin{aligned} \delta_{\mathrm{BRST}} \hat{A}_{\mu\, a} &= (\hat{D}_\mu \hat{\omega})_a \widehat{\delta\xi}, \\ \delta_{\mathrm{BRST}} \hat{\omega}_a &= -\tfrac{1}{2}\hat{g} c_{abc} \hat{\omega}_b \hat{\omega}_c \widehat{\delta\xi}, \\ \delta_{\mathrm{BRST}} \hat{\eta}_a &= \hat{\lambda}\partial \cdot \hat{A}_a \widehat{\delta\xi}. \end{aligned} \tag{3.7}$$

Here, $\widehat{\delta\xi}$ is a constant parameter with Grassmann parity 1, that is it anticommutes with the (anti)ghost field components (and the fermion field components, if there were any), but commutes with everything else. We introduce the notation

$$\frac{\delta^r_{\mathrm{BRST}}}{\widehat{\delta\xi}} \tag{3.8}$$

in analogy with the right derivatives for Grassmann variables to mean that $\widehat{\delta\xi}$ is commuted or anticommuted to the extreme right and then removed. This variation, called the BRST variation, is a symmetry of the Lagrangian density, Eq. (3.1), since the change in the Lagrangian density is a four-divergence, without invoking the equations of motion

$$\frac{\delta^r_{\mathrm{BRST}}}{\widehat{\delta\xi}} \mathcal{L} = -\hat{\lambda}\partial^\mu [(\hat{D}_\mu \hat{\omega})_a \partial \cdot \hat{A}_a]. \tag{3.9}$$

The important property of nilpotence,

$$\delta^2_{\mathrm{BRST}}(Anything) = 0, \tag{3.10}$$

holds only after using one of the equations of motion, Eq. (3.6c), which will be called the 'trivial equation of motion' in what follows.[8]

### 3.4 Energy-Momentum Tensor

The symmetric, conserved energy-momentum stress tensor density can be constructed from the canonical tensor by using Belinfante's procedure [20, 21]. This is also the tensor proven by Freedman, Muzinich, and Weinberg [14, 15] to have finite Green functions with renormalized external fields. It is

$$\begin{aligned}\theta_{\mu\nu} =& -g_{\mu\nu}\mathcal{L} - \hat{F}_{\mu\rho\ a}\hat{F}_{\nu\ a}^{\ \rho} \\ & -g_{\mu\nu}\hat{\lambda}\partial^\rho(\hat{A}_{\rho\ a}\partial\cdot\hat{A}_a) + \hat{\lambda}(\hat{A}_{\mu\ a}\partial_\nu\partial\cdot\hat{A}_a) + \hat{\lambda}(\hat{A}_{\nu\ a}\partial_\mu\partial\cdot\hat{A}_a) \\ & +(\partial_\mu\hat{\eta}_a)(\hat{D}_\nu\hat{\omega})_a + (\partial_\nu\hat{\eta}_a)(\hat{D}_\mu\hat{\omega})_a = \theta_{\nu\mu}.\end{aligned} \tag{3.11}$$

The gauge-invariant piece is

$$\theta^{(GI)}_{\mu\nu} = \tfrac{1}{4}g_{\mu\nu}\hat{F}^{\rho\pi}_a\hat{F}_{\rho\pi\ a} - \hat{F}_{\mu\rho\ a}\hat{F}_{\nu\ a}^{\ \rho}. \tag{3.12}$$

[8] If the Lagrangian formulation with the Nakanishi-Lautrup field is used, as in [18] and [19], no equations of motion are needed to demonstrate the nilpotence of the BRST variation.

The gauge-variant piece is everything else,

$$
\begin{aligned}
\theta^{(GV)}_{\mu\nu} =& \theta_{\mu\nu} - \theta^{(GI)}_{\mu\nu} \\
=& -g_{\mu\nu}[\hat{\lambda}\partial^{\rho}(\hat{A}_{\rho\ a}\partial\cdot\hat{A}_a) - \tfrac{1}{2}\hat{\lambda}(\partial\cdot\hat{A}_a)(\partial\cdot\hat{A}_a) + (\partial^{\rho}\hat{\eta}_a)(\hat{D}_{\rho}\hat{\omega})_a] \\
& +\hat{\lambda}(\hat{A}_{\mu\ a}\partial_{\nu}\partial\cdot\hat{A}_a) + \hat{\lambda}(\hat{A}_{\nu\ a}\partial_{\mu}\partial\cdot\hat{A}_a) \\
& +(\partial_{\mu}\hat{\eta}_a)(\hat{D}_{\nu}\hat{\omega})_a + (\partial_{\nu}\hat{\eta}_a)(\hat{D}_{\mu}\hat{\omega})_a.
\end{aligned}
\tag{3.13}
$$

The gauge-variant piece of the energy-momentum tensor is the BRST variation of an ‘ancestor’ operator

$$
\widehat{ancestor}\left(\theta^{(GV)}_{\mu\nu}\right) = (\partial_{\nu}\hat{\eta}_a)\hat{A}_{\mu\ a} + (\partial_{\mu}\hat{\eta}_a)\hat{A}_{\nu\ a} - g_{\mu\nu}[\tfrac{1}{2}\hat{\eta}_a\partial\cdot\hat{A}_a + (\partial_{\rho}\hat{\eta}_a)\hat{A}^{\rho}_a], \tag{3.14}
$$

since

$$
\frac{\delta^r_{\mathrm{BRST}}}{\widehat{\delta\xi}}\widehat{ancestor}\left(\theta^{(GV)}_{\mu\nu}\right) = \theta^{(GV)}_{\mu\nu} - \tfrac{1}{2}g_{\mu\nu}\hat{\eta}_a\partial^{\rho}(\hat{D}_{\rho}\hat{\omega})_a \tag{3.15}
$$

and the last term vanishes by the trivial equation of motion, Eq. (3.6c). The ancestor operator defined above is not finite, that is it does not have finite Green functions with all renormalized fields. In Section 4.4 we present the finite ancestor operator after introducing the ‘renormalized BRST variation’.

The BRST variation of the GV piece vanishes (up to the trivial equation of motion) because of the nilpotence of the BRST transformation. We say that a gauge-variant operator such as $\theta^{(GV)}_{\mu\nu}$ is BRST-exact if it has an ancestor. The BRST variation of the GI piece vanishes without using the equations of motion because the BRST variation of the gluon field is based on the gauge transformation of that same field, so any gauge-invariant quantity is automatically BRST-invariant.

### 3.5 Renormalization

We use multiplicative renormalization

$$
\begin{aligned}
\hat{A}_{\mu\ a} &= Z_A^{\frac{1}{2}} A_{\mu\ a} \\
\hat{\omega}_a &= Z_\omega^{\frac{1}{2}} \omega_a \\
\hat{\eta}_a &= Z_\eta^{\frac{1}{2}} \eta_a \\
\hat{\lambda} &= Z_\lambda \lambda \qquad \text{(It is known that } Z_\lambda = Z_A^{-1} \text{ to all orders)} \\
\hat{g} &= Z_g g \mu^\epsilon
\end{aligned}
\tag{3.16}
$$

and dimensional regularization in $4 - 2\epsilon$ space-time dimensions with the Modified Minimal Subtraction ($\overline{\mathrm{MS}}$) scheme.[9] See [13] for a thorough treatment of the subject.

The Lagrangian density, Eq. (3.1), can be written in terms of renormalized fields and parameters.

[9] Hamberg and van Neerven work in $4 + \epsilon$ dimensions.

This is the *same* Lagrangian density so the same symbol, $\mathcal{L}$, is used to represent both quantities[10]

$$
\begin{aligned}
\mathcal{L} =& -\tfrac{1}{4}F^{\mu\nu}_a F_{\mu\nu\ a} - \tfrac{1}{2}\lambda(\partial\cdot A_a)(\partial\cdot A_a) + (\partial^\mu \eta_a)(D_\mu \omega)_a \\
& -\tfrac{1}{4}\delta Z_{II}(\partial_\mu A_{\nu\ a} - \partial_\nu A_{\mu\ a})(\partial^\mu A^\nu_a - \partial^\nu A^\mu_a) \\
& +\tfrac{1}{2}\delta Z_{III} g\mu^\epsilon c_{abc}(\partial_\mu A_{\nu\ a} - \partial_\nu A_{\mu\ a})A^\mu_b A^\nu_c \\
& -\tfrac{1}{4}\delta Z_{IV} g^2\mu^{2\epsilon} c_{abc}c_{ade} A_{\mu\ b}A_{\nu\ c}A^\mu_d A^\nu_e \\
& -\tfrac{1}{2}(Z_\lambda Z_A - 1)\lambda(\partial\cdot A_a)(\partial\cdot A_a) \\
& +\delta Z_0(\partial_\mu \eta_a)\partial^\mu \omega_a \\
& -\delta Z_I g\mu^\epsilon c_{abc}(\partial_\mu \eta_a)A^\mu_b \omega_c,
\end{aligned}
\tag{3.17}
$$

where

$$
\begin{aligned}
Z_0 &\equiv Z_\omega^{\frac{1}{2}} Z_\eta^{\frac{1}{2}} \\
Z_I &\equiv Z_A^{\frac{1}{2}} Z_\omega^{\frac{1}{2}} Z_\eta^{\frac{1}{2}} Z_g \\
Z_{II} &\equiv Z_A \\
Z_{III} &\equiv Z_A^{\frac{3}{2}} Z_g \\
Z_{IV} &\equiv Z_A^2 Z_g^2
\end{aligned}
\tag{3.18}
$$

Notice that the renormalized coupling $g$ is dimensionless and that we have introduced a parameter $\mu$ with the dimensions of mass. The renormalization constants $\delta Z_i \equiv Z_i - 1$ have been given Roman subscripts which label the number of gauge fields in the counterterm vertex.

Since the coupling is universal, the three different interaction vertices have associated renor-

[10] Some authors use the term 'renormalized Lagrangian density,' but it is not always clear what is meant.

malization constants related by the following 'renormalization constant Ward identities' :

$$Z_g = Z_I Z_A^{-\frac{1}{2}} Z_0^{-1} = Z_{III} Z_A^{-\frac{3}{2}} = Z_{IV}^{\frac{1}{2}} Z_A^{-1} \tag{3.19}$$

In the Minimal Subtraction (MS) renormalization scheme, the counterterms are the negative of the pole part only, with no finite component. In the Modified Minimal Subtraction ($\overline{\text{MS}}$) scheme, the ubiquitous Euler's constant, $\gamma_E$, and the natural logarithms of $4\pi$ are absorbed into a new renormalization mass parameter, $\bar{\mu}$, defined by

$$\bar{\mu} \equiv \mu \left(\frac{4\pi}{e^{\gamma_E}}\right)^{\frac{1}{2}}. \tag{3.20}$$

Applied under dimensional regularization, the counterterms in either scheme are proportional to $\frac{1}{\epsilon}$.

We list the multiplicative renormalization constants in the $\overline{\text{MS}}$ scheme to the order needed for an $\mathcal{O}(g^2)$ calculation of the Green functions considered in this paper

$$\begin{aligned}
Z_0 &= 1 + \frac{1}{\epsilon}\frac{g^2}{16\pi^2} C_A \left[\frac{1}{2} + \frac{1}{4}\left(1 - \frac{1}{\lambda}\right)\right] + \mathcal{O}(g^4) \\
Z_A &= 1 + \frac{1}{\epsilon}\frac{g^2}{16\pi^2} C_A \left[\frac{5}{3} + \frac{1}{2}\left(1 - \frac{1}{\lambda}\right)\right] + \mathcal{O}(g^4) \\
Z_\lambda &= 1 - \frac{1}{\epsilon}\frac{g^2}{16\pi^2} C_A \left[\frac{5}{3} + \frac{1}{2}\left(1 - \frac{1}{\lambda}\right)\right] + \mathcal{O}(g^4) = Z_A^{-1} \\
Z_g &= 1 + \mathcal{O}(g^2)
\end{aligned} \tag{3.21}$$

where $C(A) = N$ for the gauge group $SU(N)$.

## 3.6 (Modified) LSZ Reduction

The residue of the propagator pole is used in the LSZ (for Lehmann, Symanzik, and Zimmermann) reduction formula to derive the S-matrix from Green functions. The basic idea is that the S-matrix is obtained from the asymptotic behavior of Green functions for large times ($t \to \pm\infty$), and this behavior is governed by the singularities of the external propagators. We use a modified version of this procedure to handle the infra-red divergent logarithms that appear in this massless theory.[11]

Let $\Sigma(p)$ be the self-energy. Then

$\Sigma^{\sigma\tau}_{gluon}(p)\delta_{ab} \equiv i \Big($ 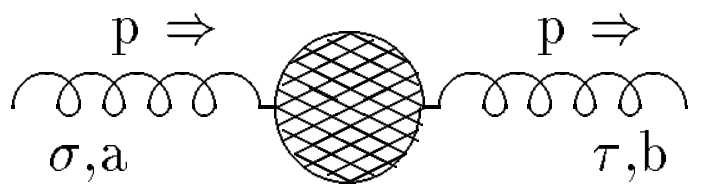

$\Big)$, (3.22a)

$\Sigma_{ghost}(p)\delta_{ab} \equiv i \Big($ 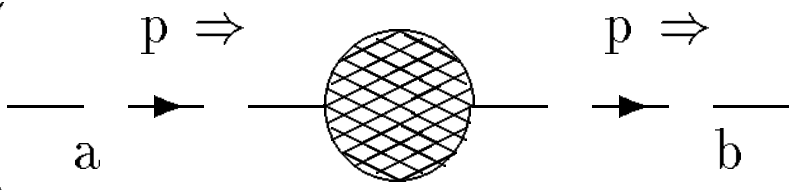

$\Big)$, (3.22b)

where the cross-hatched blob represents all one-particle irreducible amputated graphs, including counterterms so that the blob has no ultra-violet (UV) divergences. We isolate the $p^2$ dependence, defining $\Pi(p^2)$ by

$$\Sigma^{\sigma\tau}_{gluon}(p) = (p^2 g^{\sigma\tau} - p^\sigma p^\tau)\Pi(p^2), \tag{3.23}$$

noting that the gluon self-energy is purely transverse due to a Ward identity [13].

In a massless theory, the singularities in propagators, as a function of $p^2$, are not simple poles after higher order corrections are included.

The dressed propagators are then

$$\begin{aligned}\mathcal{D}^{\sigma\tau}_{ab}(p) &\equiv \frac{i\delta_{ab}}{p^2+i\epsilon}\left\{-g^{\sigma\tau}\frac{1}{1+\Pi(p^2)} + \frac{p^\sigma p^\tau}{p^2+i\epsilon}\left[\left(1-\tfrac{1}{\lambda}\right) - \frac{\Pi(p^2)}{1+\Pi(p^2)}\right]\right\} \\ &\xrightarrow[p^2\to 0]{} \frac{-i\delta_{ab}c^2_{gluon}g^{\sigma\tau}}{p^2+i\epsilon} + \frac{i\delta_{ab}\tilde{c}^2_{gluon}p^\sigma p^\tau}{(p^2+i\epsilon)^2},\end{aligned} \tag{3.24a}$$

and

$$\mathcal{D}_{ab}(p) \equiv \frac{i\delta_{ab}}{p^2 - \Sigma_{ghost}(p) + i\epsilon} \xrightarrow[p^2\to 0]{} \frac{i\delta_{ab}c^2_{ghost}}{p^2+i\epsilon}, \tag{3.24b}$$

where $c^2$ is the residue of the propagator pole (the coefficient of $p^2$ in the denominator). To one-loop

[11] We do not have a complete justification of our algorithm.

order, the gluon self-energy is

$$\Sigma^{\sigma\tau}_{gluon}(p) = \frac{g^2}{16\pi^2} C_A \left(p^2 g_{\sigma\tau} - p_\sigma p_\tau\right) \left\{ \begin{array}{l} -\frac{1}{4}\left(1-\frac{1}{\lambda}\right)^2 \\ +\left(1-\frac{1}{\lambda}\right)\left[1+\frac{1}{2}\ln\left(\frac{-p^2}{\bar{\mu}^2}\right)\right] \\ -\frac{31}{9}+\frac{5}{3}\ln\left(\frac{-p^2}{\bar{\mu}^2}\right) \end{array} \right\} + \mathcal{O}(g^4), \tag{3.25}$$

which implies that

$$\Pi(p^2) = \frac{g^2}{16\pi^2} C_A \left\{ \begin{array}{l} -\frac{1}{4}\left(1-\frac{1}{\lambda}\right)^2 \\ +\left(1-\frac{1}{\lambda}\right)\left[1+\frac{1}{2}\ln\left(\frac{-p^2}{\bar{\mu}^2}\right)\right] \\ -\frac{31}{9}+\frac{5}{3}\ln\left(\frac{-p^2}{\bar{\mu}^2}\right) \end{array} \right\} + \mathcal{O}(g^4), \tag{3.26}$$

and the ghost self-energy to one-loop order is

$$\Sigma_{ghost}(p) = \frac{g^2}{16\pi^2} C_A p^2 \left[1 - \frac{1}{2}\ln\left(\frac{-p^2}{\bar{\mu}^2}\right) - \frac{1}{4}\left(1-\frac{1}{\lambda}\right)\ln\left(\frac{-p^2}{\bar{\mu}^2}\right)\right] + \mathcal{O}(g^4). \tag{3.27}$$

The singularities in the propagators are not simple poles, but the leading power, with logarithmic corrections, is governed by the large-time behavior of the propagator. So to define the residue we use the following formulae, which would be valid when the physical masses are nonzero:

$$\begin{aligned} c^2_{gluon} &= \left[ \frac{\partial}{\partial p^2}\{p^2[1+\Pi(p^2)]\}\Big|_{p^2=m^2_{physical}} \right]^{-1} \\ &= 1 - \tfrac{g^2}{16\pi^2} C_A \left\{ -\tfrac{1}{4}\left(1-\tfrac{1}{\lambda}\right)^2 + \left(1-\tfrac{1}{\lambda}\right)\left[\tfrac{3}{2}+\tfrac{1}{2}\ln\left(\tfrac{-p^2}{\bar{\mu}^2}\right)\right] - \tfrac{16}{9} + \tfrac{5}{3}\ln\left(\tfrac{-p^2}{\bar{\mu}^2}\right) \right\}\Big|_{p^2=0} + \mathcal{O}(g^4), \end{aligned} \tag{3.28a}$$

and

$$
\begin{aligned}
c^2_{ghost} &= \left[ 1 - \left. \frac{\partial \Sigma_{ghost}(p)}{\partial p^2} \right|_{p^2 = m^2_{physical}} \right]^{-1} \\
&= 1 - \tfrac{g^2}{16\pi^2} C_A \left\{ \left(1 - \tfrac{1}{\lambda}\right) \left[ \tfrac{1}{4} + \tfrac{1}{4} \ln\left(\tfrac{-p^2}{\bar{\mu}^2}\right) \right] - \tfrac{1}{2} + \tfrac{1}{2} \ln\left(\tfrac{-p^2}{\bar{\mu}^2}\right) \right\} \Big|_{p^2=0} + \mathcal{O}(g^4).
\end{aligned}
\qquad (3.28b)
$$

We do not calculate $\tilde{c}^2_{gluon}$ because it is not used in this paper. It is important to refrain from taking the on-shell limit ($p^2 \to 0$) until the IR divergent logarithms above have cancelled algebraically with similar logarithms in the amputated Green functions which are being converted to S-matrix elements.

We have generalized the notion of residue to include the IR divergent terms that arise in a massless theory. Notice that we are extracting the residue of the propagator pole by taking the partial derivative of the denominator with respect to $p^2$ at $p^2 = 0$, rather than simply dividing the denominator by $p^2$. The partial derivative extracts that piece of the IR-divergent logarithm which is proportional to $p^2$. This piece is necessary to ensure, for example, that the two-gluon matrix element of the energy-momentum tensor, Eq. (4.2), is IR-finite on shell and equal to its correct value.

### 3.7 Covariant Gluon Operator

In [8], Hamberg and van Neerven calculate the anomalous dimension of the covariant gluon operator to two-loop order, $\mathcal{O}(\alpha_S^2)$, with all free Lorentz indices contracted with a null-vector $\Delta$. This selects the highest-spin part of the operator and eliminates the need to calculate the trace terms.

The covariant gluon operator is

$$
\mathcal{O}_g^{\mu_1 \cdots \mu_m}(x) = \tfrac{1}{2} i^{m-2} \mathcal{S}[\hat{F}^{\rho\mu_1}_{c_1}(x) \hat{D}^{\mu_2}_{c_1 c_2}(x) \hat{D}^{\mu_3}_{c_2 c_3}(x) \ldots \hat{D}^{\mu_{m-1}}_{c_{m-1} c_m}(x) \hat{F}_{\rho}{}^{\mu_m}_{c_m}(x)] + \text{trace terms}, \qquad (3.29)
$$

where $\mathcal{S}$ denotes symmetrization of the Lorentz indices $\mu_i$ and the trace terms make the operator traceless under all possible contractions of the free Lorentz indices in pairs. The $c_i$ are color indices in the adjoint representation.

Hamberg and van Neerven's gauge-invariant operator is

$$\mathcal{O}_g^{(m)} = \mathcal{O}^{\mu_1 \cdots \mu_m} \Delta_{\mu_1} \ldots \Delta_{\mu_m} \quad , \tag{3.30}$$

where $\Delta$ is light-like. In units such that $c = 1 = \hbar$, the mass-dimension of this operator grows linearly with $m$,

$$[\mathcal{O}_g^{(m)}] = m + 2 \tag{3.31}$$

Selecting the highest-spin piece is equivalent to selecting the lowest twist, since

$$\text{twist} \;\equiv\; \text{mass-dimension} \;-\; \text{spin}. \tag{3.32}$$

All operators of the form above, Eq. (3.30), are twist-2.

The simplest case ($m = 2$) of the covariant gluon operator, Eq. (3.29), gives, up to a multiplicative factor, the gauge-invariant part of the energy-momentum tensor, Eq. (3.12)

$$\mathcal{O}_g^{\mu\nu} = \tfrac{1}{2} \hat{F}_a^{\mu\rho} \hat{F}^{\nu}_{\rho\ a} - \tfrac{1}{8} g^{\mu\nu} \hat{F}_a^{\rho\pi} \hat{F}_{\rho\pi\ a} = -\tfrac{1}{2} \theta^{(GI)\ \mu\nu}. \tag{3.33}$$

We study this case because of the relative simplicity of the calculation, but also because the gauge-variant operators which mix with it are supposed to be known [14, 15]; they are the gauge-variant operators in the energy-momentum tensor, Eq. (3.13). No other operators are required. The gauge-variant operators for our special case also happen to be those given by the Joglekar–Lee prescription. We calculate the trace terms mentioned in Eq. (3.29) explicitly even though they are higher-twist.

The specific case $m = 2$ of Hamberg and van Neerven's operator, Eq. (3.30), is

$$\mathcal{O}_g^{(2)} = \mathcal{O}_g^{\mu\nu} \Delta_\mu \Delta_\nu = \tfrac{1}{2} \hat{F}^{\mu}_{\ \rho\ a} \hat{F}_a^{\nu\rho} \Delta_\mu \Delta_\nu. \tag{3.34}$$

Hamberg and van Neerven use a basis of operators given by Dixon and Taylor [22] before the BRST symmetry was fully developed. For the case $m = 2$, their alien operator does not correspond to our GV operator (Eq 3.13). The basis of operators that they chose to mix with their GI operator, Eq. (3.34), is

$$\mathcal{O}^{(2)}_{alien} = (\hat{F}^{\rho\mu}_a \hat{D}_{\rho\ ab} \hat{A}^{\nu}_b + \hat{\eta}_a \partial^{\mu}\partial^{\nu}\hat{\omega}_a)\Delta_\mu\Delta_\nu + \mathcal{O}(g^2), \tag{3.35}$$

where

$$\begin{aligned}\mathcal{O}^{\mu\nu}_{alien} =& \tfrac{1}{2}[\hat{F}^{\rho\mu}_a \hat{D}_{\rho\ ab}\hat{A}^{\nu}_b + \hat{F}^{\rho\nu}_a \hat{D}_{\rho\ ab}\hat{A}^{\mu}_b] + \hat{\eta}_a\partial^\mu\partial^\nu\hat{\omega}_a + \text{terms proportional to } g^{\mu\nu} + \mathcal{O}(\hat{g}^2) \\ =& \tfrac{1}{2}[\hat{F}^{\rho\mu}_a (\hat{D}_\rho \hat{A}^{\nu})_a + \hat{F}^{\rho\nu}_a (\hat{D}_\rho \hat{A}^{\mu})_a] + \hat{\eta}_a\partial^\mu\partial^\nu\hat{\omega}_a + \text{terms proportional to } g^{\mu\nu} + \mathcal{O}(\hat{g}^2).\end{aligned} \tag{3.36}$$

This operator is not BRST-exact, in fact its BRST variation does not vanish.

We can give a schematic form for the operators which are BRST ancestors to the alien operators of highest twist that mix with the GI operators for all even moments $m$ simply by counting mass-dimension and twist, and by requiring an $SU(N)$ singlet.

$$\widehat{ancestor}\ (\mathcal{O}^{\mu_1\cdots\mu_m}_{alien}) = \sum_{i=0}^{\frac{m}{2}-1} \mathcal{C}^{(m)}_i \ \ \mathcal{S}\left[(\hat{\eta}\cdot\hat{A}^{\mu_1}) \ \prod_{j=1}^{i}(\hat{A}^{\mu_{2j}}\cdot\hat{A}^{\mu_{2j+1}}) \ \prod_{k=1}^{m-2i-1}(\partial^{\mu_{m-k+1}})\right] + \mathcal{O}(g), \tag{3.37}$$

where $\mathcal{S}$ denotes symmetrization of the Lorentz indices and the coefficients $\mathcal{C}^{(m)}_i$ are arbitrary (they are determined as a result of the renormalization). The derivatives may act on any combination of the fields. The dot products represent contractions over color indices. The alien operators obtained from these ancestors, of course, have vanishing BRST variation (modulo the trivial equation of motion), because of the nilpotence of that variation.

Some examples should clarify the notation. The ancestor of the twist-2 GV part of energy-momentum tensor (second moment of the covariant gluon operator) in schematic form is

$$\widehat{ancestor}\ (\mathcal{O}^{\mu_1\mu_2}_{alien}) = C^{(2)}_0[\hat{\eta}_a\hat{A}^{\mu_1}_a\partial^{\mu_2} + \hat{\eta}_a\hat{A}^{\mu_2}_a\partial^{\mu_1}] + \mathcal{O}(g). \tag{3.38}$$

Compare this with the twist-2 part of Eq. (3.14). The fields upon which the partial derivatives act are not specified in the schematic form. The arbitrary constant $C_0^{(2)}$ can be absorbed into entries of the renormalization mixing matrix. The $\mathcal{O}(g)$ terms turn out to be unnecessary.

The dimension-6 GI operator (fourth moment of the covariant gluon operator) is

$$\mathcal{O}_g^{\mu_1\mu_2\mu_3\mu_4} = -\frac{1}{2}\mathcal{S}[\hat{F}^{\rho\mu_1}_{c_1}\hat{D}^{\mu_2}_{c_1c_2}\hat{D}^{\mu_3}_{c_2c_3}\hat{F}_{\rho\ c_3}^{\ \mu_4}] + \text{trace terms.} \tag{3.39}$$

The BRST ancestor of the GV operators that mix with the twist-2 (highest-twist) part of the GI operator above is

$$\widehat{ancestor}\,(\mathcal{O}_{alien}^{\mu_1\mu_2\mu_3\mu_4}) = C_0^{(4)}\mathcal{S}[\hat{\eta}_a\hat{A}_a^{\mu_1}\partial^{\mu_2}\partial^{\mu_3}\partial^{\mu_4}] + C_1^{(4)}\mathcal{S}[\hat{\eta}_a\hat{A}_a^{\mu_1}\hat{A}_b^{\mu_2}\hat{A}_b^{\mu_3}\partial^{\mu_4}] + \mathcal{O}(g). \tag{3.40}$$

Again, the partial derivatives may act on any combination of the fields. Only one of the two arbitrary constants may be absorbed into the mixing matrix; the ratio $\frac{C_0^{(4)}}{C_1^{(4)}}$ is determined by the renormalization of $\mathcal{O}_g^{\mu_1\mu_2\mu_3\mu_4}$.

# 4 Results of the Calculation at Zero Momentum Transfer

The results in this section were obtained under the possibly questionable operation[12] of first taking the limit of zero momentum transfer, before any of the Feynman diagrams are evaluated. The software package used was the symbolic manipulator, FORM [23], written by J.A.M. Vermaseren. We now evaluate Green functions of various pieces of the energy-momentum tensor first with two gluon fields, then with one ghost field and one anti-ghost field.

The fields in the inserted operators are bare, while the external gluon and ghost legs are renormalized as usual, since it is the renormalized fields that interact and are loop-corrected. This differs from Hamberg and van Neerven's calculation in which the external fields as well as the fields in the inserted operator are bare. The two calculations therefore differ by factors of some multiplicative renormalization constants, Eq. (3.21).

[12]One might be suspicious of interchanging the order of the limit as the momentum transfer goes to zero with the other limits involved in the renormalization procedure.

## 4.1 Green Functions of $\theta_{\mu\nu}$ with Two Gluon Fields

While we give the pole pieces in their entirety, the finite parts have been simplified for clarity in presentation. The full tensor structure can be found in the Appendix, where we also list the twist-2 (spin-2) piece of the operators.

### 4.1.1 Entire Energy-Momentum Tensor

Consider the amputated gluon two-point Green function with the entire energy-momentum tensor, Eq. (3.11), inserted at zero momentum transfer. The external gluon fields have not been contracted with physical polarization vectors, we have not multiplied by the modified LSZ residue of the gluon propagator pole, and the external momenta have not been put on shell. This is what we will mean by an off-shell gluon Green function in the sections that follow. Explicit calculation gives

$$\langle 0|TA_{\sigma\ a}\theta_{\mu\nu}A_{\tau\ b}|0\rangle_{Amputated} = p_\mu p_\nu g_{\sigma\tau}\delta_{ab}\left(-2+\frac{g^2}{16\pi^2}C_A\left\{\begin{array}{l}\frac{1}{2}\left(1-\frac{1}{\lambda}\right)^2\\ -\left(1-\frac{1}{\lambda}\right)\left[3+\ln\left(\frac{-p^2}{\bar{\mu}^2}\right)\right]\\ +\frac{32}{9}-\frac{10}{3}\ln\left(\frac{-p^2}{\bar{\mu}^2}\right)\end{array}\right\}\right) \tag{4.1}$$

$$+\text{UV-finite terms that vanish on shell} + \mathcal{O}(g^4).$$

Notice that this object is UV-finite. A glance at Eq. (A.2) in the Appendix will satisfy the reader that the terms not included above are also UV-finite, even off shell. The UV-finiteness supports the results of Freedman *et al.* [14,15] on the renormalization of the symmetric energy-momentum tensor. Also, since there is no pole, the contribution to the anomalous dimension is zero.

Taking into account Hamberg and van Neerven's different dimensional regularization prescrip-tion (dimension $4+\epsilon$ instead of $4-2\epsilon$) and their use of bare fields instead of renormalized fields for the external legs, our result contracted with light-like vectors, Eq. (A.3), agrees with theirs.

We now do what Hamberg and van Neerven argue, quite reasonably, in their paper is impossible. We construct a matrix element between massless gluon states, but to do so we use the modified LSZ prescription described in Section 3.6.

Contracting with physical polarization vectors, using the modified LSZ residue of the gluon propagator pole, $c^2_{gluon}$ in Eq. (3.28a), and putting the external momenta on shell, we get the relatively simple S-matrix element

$$\langle \epsilon_1, p, a | \theta_{\mu\nu} | \epsilon_2, p, b \rangle = -2 p_\mu p_\nu \delta_{ab} \epsilon_1^* \cdot \epsilon_2 + \mathcal{O}(g^4), \tag{4.2}$$

where we have used the fact that $\epsilon_i$ is a physical polarization vector satisfying

$$p \cdot \epsilon_i = 0 \qquad\qquad i = 1, 2. \tag{4.3}$$

The physical state $|\epsilon_i, p, a\rangle$ is meant to represent an on-shell gluon of momentum $p$, polarization vector $\epsilon_i$, and color $a$.

The modified LSZ procedure eliminates the IR divergent logarithms even before the external momenta are taken on shell. The result Eq. (4.2) is not surprising since $\theta_{\mu\nu}$ is the conserved Noether current and

$$P_\nu \equiv \int d^3\mathbf{x}\ \theta_{0\nu}(x) \tag{4.4}$$

is the Noether charge. It measures the physical (non-IR-divergent) energy-momentum in a state. A correct calculation should show that all the higher-order corrections to the right-hand side of Eq. (4.2) vanish.

In the next two sections, we examine the GI and GV pieces separately.

### 4.1.2 Gauge-Invariant Part

Consider now the amputated off-shell gluon two-point Green function with only the gauge-invariant piece of the energy-momentum tensor, Eq. (3.12), inserted at zero momentum transfer. Explicit calculation gives

$$\begin{aligned}
\langle 0|TA_{\sigma\ a}\theta^{(GI)}_{\mu\nu}A_{\tau\ b}|0\rangle_{Amputated} =& \frac{1}{\epsilon}\frac{g^2}{16\pi^2}C_A\delta_{ab}\left\{\begin{array}{l} p^2(-\frac{1}{2}g_{\sigma\tau}g_{\mu\nu}+g_{\sigma\mu}g_{\tau\nu}+g_{\sigma\nu}g_{\tau\mu}) \\ \\ -\frac{1}{2}(p_\tau p_\nu g_{\sigma\mu}+p_\tau p_\mu g_{\sigma\nu}+p_\sigma p_\nu g_{\tau\mu}+p_\sigma p_\mu g_{\tau\nu}-p_\sigma p_\tau g_{\mu\nu}) \end{array}\right\} \\
&+p_\mu p_\nu g_{\sigma\tau}\delta_{ab}\left(-2+\frac{g^2}{16\pi^2}C_A\left\{\begin{array}{l} \left(1-\frac{1}{\lambda}\right)^2 \\ -\left(1-\frac{1}{\lambda}\right)\left[6+\ln\left(\frac{-p^2}{\bar{\mu}^2}\right)\right] \\ +\frac{86}{9}-\frac{10}{3}\ln\left(\frac{-p^2}{\bar{\mu}^2}\right) \end{array}\right\}\right) \\
&+\text{UV-finite terms that vanish on shell}+\mathcal{O}(g^4).
\end{aligned} \tag{4.5}$$

There is a UV pole in Eq. (4.5), but this UV divergence vanishes on shell with physical polarizations.

A further clue that we have performed the calculation correctly is the fact that Eq. (A.4) satisfies the following Ward identity

$$p_\sigma p_\tau\langle 0|TA_{\sigma\ a}\theta_{\mu\nu}A_{\tau\ b}|0\rangle_{Amputated}=0. \tag{4.6}$$

If we now put this result on mass-shell and use the modified LSZ procedure to derive the S-matrix element, we get

$$\langle\epsilon_1,p,a|\theta^{(GI)}_{\mu\nu}|\epsilon_2,p,b\rangle = p_\mu p_\nu\delta_{ab}\epsilon_1^*\cdot\epsilon_2\left\{-2+\frac{g^2}{16\pi^2}C_A\left[\frac{1}{2}\left(1-\frac{1}{\lambda}\right)^2-3\left(1-\frac{1}{\lambda}\right)+6\right]\right\}+\mathcal{O}(g^4). \tag{4.7}$$

Notice that this physical matrix element of a gauge-invariant operator depends on the gauge-fixing parameter, $\lambda$. Also, the GI part is not equal to the total, so we must calculate the GV part.

### 4.1.3 Gauge-Variant (Alien) Part

Consider the amputated off-shell gluon two-point Green function with only the gauge-variant piece of the energy-momentum tensor, Eq. (3.13), inserted at zero momentum transfer. Explicit calculation gives

$$
\begin{aligned}
\langle 0|TA_{\sigma\, a}\theta^{(GV)}_{\mu\nu}A_{\tau\, b}|0\rangle_{Amputated} =& \frac{1}{\epsilon}\frac{g^2}{16\pi^2}C_A\delta_{ab}\left\{\begin{array}{l} p^2(\frac{1}{2}g_{\sigma\tau}g_{\mu\nu}-g_{\sigma\mu}g_{\tau\nu}-g_{\sigma\nu}g_{\tau\mu}) \\ \\ +\frac{1}{2}(p_\tau p_\nu g_{\sigma\mu}+p_\tau p_\mu g_{\sigma\nu}+p_\sigma p_\nu g_{\tau\mu}+p_\sigma p_\mu g_{\tau\nu}-p_\sigma p_\tau g_{\mu\nu})\end{array}\right\} \\
&+p_\mu p_\nu g_{\sigma\tau}\delta_{ab}\frac{g^2}{16\pi^2}C_A\left\{-\frac{1}{2}\left(1-\frac{1}{\lambda}\right)^2+3\left(1-\frac{1}{\lambda}\right)-6\right\} \\
&+\text{UV-finite terms that vanish on shell}+\mathcal{O}(g^4).
\end{aligned}
\tag{4.8}
$$

The pole terms cancel between the GI and GV parts off shell. On shell, each pole piece vanishes individually.

Going on-shell and using the modified LSZ procedure, we get the S-matrix element

$$
\langle\epsilon_1,p,a|\theta^{(GV)}_{\mu\nu}|\epsilon_2,p,b\rangle = p_\mu p_\nu\delta_{ab}\epsilon_1^*\cdot\epsilon_2\frac{g^2}{16\pi^2}C_A\left[-\frac{1}{2}\left(1-\frac{1}{\lambda}\right)^2+3\left(1-\frac{1}{\lambda}\right)-6\right]+\mathcal{O}(g^4). \tag{4.9}
$$

Notice that the finite part of the physical matrix element does not vanish. But, $\theta^{(GV)}_{\mu\nu}$ is BRST-exact, as we observed, and there is a general theorem that BRST-exact operators have vanishing physical matrix elements. Thus, we know that we have a contradiction with general theorems.

Hamberg and van Neerven have a similar result implicit in their formulae (they did not remark on it), but since their alien operators are not BRST-exact, physical matrix elements of their alien operators would not be expected to vanish.

## 4.2 Green Functions of $\theta_{\mu\nu}$ with One Ghost Field and One Anti-Ghost Field

### 4.2.1 Entire Energy-Momentum Tensor

We have just seen that the two-gluon matrix element of the gauge-variant part of the energy-momentum tensor is nonzero. Since $\theta^{(GV)}_{\mu\nu}$ is BRST-exact, this contradicts a crucial part of the theory on the renormalization of gauge-invariant operators and so we cannot take for granted any of the results of this theory, but must verify the results.

In particular, we need to verify the finiteness at one-loop order of Green functions of the energy-momentum tensor. The previous section has established this for the gluon two-point Green function and in this section we verify finiteness for the ghost–anti-ghost Green function.

All the necessary counterterms are determined by the formula for $\theta_{\mu\nu}$; they are obtained by expanding $\theta_{\mu\nu}$ in terms of renormalized fields by Eq. (3.16).

Consider the amputated off-shell ghost two-point Green function with the entire energy-momentum tensor, Eq. (3.11), inserted at zero momentum transfer. Explicit calculation gives

$$\begin{aligned}\langle 0|T\omega_a\theta_{\mu\nu}\eta_b|0\rangle_{Amputated} =& \delta_{ab}(2p_\mu p_\nu - p^2 g_{\mu\nu}) \\ &+\frac{g^2}{16\pi^2}C_A\delta_{ab}\left\{\begin{array}{l}\left(1-\frac{1}{\lambda}\right)\left\{\frac{1}{2}\left[1+\ln\left(\frac{-p^2}{\bar{\mu}^2}\right)\right]p_\mu p_\nu - \frac{1}{4}\ln\left(\frac{-p^2}{\bar{\mu}^2}\right)p^2 g_{\mu\nu}\right\} \\ +\left[-1+\ln\left(\frac{-p^2}{\bar{\mu}^2}\right)\right]p_\mu p_\nu + \left[1-\frac{1}{2}\ln\left(\frac{-p^2}{\bar{\mu}^2}\right)\right]p^2 g_{\mu\nu}\end{array}\right\} \\ &+\mathcal{O}(g^4).\end{aligned} \tag{4.10}$$

Using the modified LSZ residue of the ghost propagator pole, $c^2_{ghost}$ in Eq. (3.28b), and putting the external momenta on shell, we get the comparatively simple S-matrix element

$$\langle p, a|\theta_{\mu\nu}|p, b\rangle = 2p_\mu p_\nu\delta_{ab} + \mathcal{O}(g^4), \tag{4.11}$$

which is correct for the expectation value of $\theta_{\mu\nu}$ in a properly normalized state of momentum $p$.

Here the state vector $|p,a\rangle$ is meant to represent an on-shell ghost of momentum $p$ and color $a$. Again, although we only performed the one-loop calculation, all higher-order corrections should vanish.

The twist-2 (spin-2) piece of the amputated Green function Eq. (4.10) above in which the free Lorentz indices of the inserted operator, $\mu$ and $\nu$, are contracted with a null-vector, $\Delta$, is

$$\langle 0|T\omega_a \Delta^\mu \theta_{\mu\nu}\Delta^\nu \eta_b|0\rangle_{Amputated} = (p\cdot\Delta)^2\delta_{ab}\left(2+\tfrac{g^2}{16\pi^2}C_A\left\{\tfrac{1}{2}\left(1-\tfrac{1}{\lambda}\right)\left[1+\ln\left(\frac{-p^2}{\bar{\mu}^2}\right)\right]-1+\ln\left(\frac{-p^2}{\bar{\mu}^2}\right)\right\}\right)+\mathcal{O}(g^4). \tag{4.12}$$

### 4.2.2 Gauge-Invariant Part

Consider now the amputated off-shell ghost two-point Green function with only the gauge-invariant piece of the energy-momentum tensor, Eq. (3.12), inserted at zero momentum transfer. Explicit calculation gives

$$\begin{aligned}\langle 0|T\omega_a \theta^{(GI)}_{\mu\nu}\eta_b|0\rangle_{Amputated} =& \tfrac{1}{\epsilon}\tfrac{g^2}{16\pi^2}C_A\delta_{ab}(p_\mu p_\nu-\tfrac{1}{4}p^2 g_{\mu\nu})\\ &+\tfrac{g^2}{16\pi^2}C_A\delta_{ab}\left\{\left[1-\ln\left(\frac{-p^2}{\bar{\mu}^2}\right)\right]p_\mu p_\nu+\tfrac{1}{4}\ln\left(\frac{-p^2}{\bar{\mu}^2}\right)p^2 g_{\mu\nu}\right\}+\mathcal{O}(g^4).\end{aligned} \tag{4.13}$$

The twist-2 (spin-2) piece of this amputated Green function in which the free Lorentz indices of the inserted operator, $\mu$ and $\nu$, are contracted with a null-vector, $\Delta$, is

$$\langle 0|T\omega_a \Delta^\mu \theta^{(GI)}_{\mu\nu}\Delta^\nu \eta_b|0\rangle_{Amputated} = (p\cdot\Delta)^2\frac{g^2}{16\pi^2}C_A\delta_{ab}\left[\frac{1}{\epsilon}+1-\ln\left(\frac{-p^2}{\bar{\mu}^2}\right)\right]+\mathcal{O}(g^4). \tag{4.14}$$

### 4.2.3 Gauge-Variant (Alien) Part

Consider the amputated off-shell ghost two-point Green function with only the gauge-variant piece of the energy-momentum tensor, Eq. (3.13), inserted at zero momentum transfer. Explicit calcula-

tion gives

$$
\begin{aligned}
\langle 0|T\omega_a\theta^{(GV)}_{\mu\nu}\eta_b|0\rangle_{Amputated} =& \tfrac{1}{\epsilon}\tfrac{g^2}{16\pi^2}C_A\delta_{ab}(-p_\mu p_\nu + \tfrac{1}{4}p^2 g_{\mu\nu}) \\
&+\delta_{ab}(2p_\mu p_\nu - p^2 g_{\mu\nu}) \\
&+\tfrac{g^2}{16\pi^2}C_A\delta_{ab}\left\{\begin{array}{l} \left(1-\tfrac{1}{\lambda}\right)\left\{\tfrac{1}{2}\left[1+\ln\left(\frac{-p^2}{\bar{\mu}^2}\right)\right]p_\mu p_\nu - \tfrac{1}{4}\ln\left(\frac{-p^2}{\bar{\mu}^2}\right)p^2 g_{\mu\nu}\right\} \\ +2\left[-1+\ln\left(\frac{-p^2}{\bar{\mu}^2}\right)\right]p_\mu p_\nu + \left[1-\tfrac{3}{4}\ln\left(\frac{-p^2}{\bar{\mu}^2}\right)\right]p^2 g_{\mu\nu} \end{array}\right\} \\
&+\mathcal{O}(g^4).
\end{aligned}
\tag{4.15}
$$

Notice that the pole terms cancel between the GI and GV pieces.

The twist-2 (spin-2) piece of this amputated Green function in which the free Lorentz indices of the inserted operator, $\mu$ and $\nu$, are contracted with a null-vector, $\Delta$, is

$$
\begin{aligned}
&\langle 0|T\omega_a\Delta^\mu\theta^{(GV)}_{\mu\nu}\Delta^\nu\eta_b|0\rangle_{Amputated} = \\
&\quad (p\cdot\Delta)^2\delta_{ab}\left(2+\tfrac{g^2}{16\pi^2}C_A\left\{-\tfrac{1}{\epsilon}+\tfrac{1}{2}\left(1-\tfrac{1}{\lambda}\right)\left[1+\ln\left(\frac{-p^2}{\bar{\mu}^2}\right)\right]-2+2\ln\left(\frac{-p^2}{\bar{\mu}^2}\right)\right\}\right)+\mathcal{O}(g^4).
\end{aligned}
\tag{4.16}
$$

## 4.3 Renormalization Mixing Matrix

If we do not require *a priori* that the matrix be triangular, then the most general form is

$$
\begin{pmatrix} R\left[\theta^{(GI)}_{\mu\nu}\right] \\ R\left[\theta^{(GV)}_{\mu\nu}\right] \\ R\left[E_{\mu\nu}\right] \end{pmatrix} = \begin{pmatrix} Z_{GG} & Z_{GA} & Z_{GE} \\ Z_{AG} & Z_{AA} & Z_{AE} \\ 0 & 0 & Z_{EE} \end{pmatrix} \begin{pmatrix} \theta^{(GI)}_{\mu\nu} \\ \theta^{(GV)}_{\mu\nu} \\ E_{\mu\nu} \end{pmatrix}, \tag{4.17}
$$

where the operator of class $E$, which vanishes by the equations of motion (Eqs. 3.6) and mixes with the operators in the energy-momentum tensor, is

$$\begin{aligned}
E_{\mu\nu} =& \delta_{ab}\hat{A}_{\mu\ c}[(\hat{D}_\rho \hat{F}^{\rho\nu})_c + \hat{\lambda}\partial^\nu \partial \cdot \hat{A}_c + \hat{g}c_{ade}(\partial^\nu \hat{\eta}_d)\hat{\omega}_e] \\
&+\delta_{ab}\hat{A}_{\nu\ c}[(\hat{D}_\rho \hat{F}^{\rho\mu})_c + \hat{\lambda}\partial^\mu \partial \cdot \hat{A}_c + \hat{g}c_{cde}(\partial^\mu \hat{\eta}_d)\hat{\omega}_e] \\
&-\tfrac{1}{2}\delta_{ab}g_{\mu\nu}\hat{A}_{\pi\ c}[(\hat{D}_\rho \hat{F}^{\rho\pi})_c + \hat{\lambda}\partial^\pi \partial \cdot \hat{A}_c + \hat{g}c_{cde}(\partial^\pi \hat{\eta}_d)\hat{\omega}_e] \\
&+\tfrac{1}{2}\delta_{ab}g_{\mu\nu}[(\hat{D}_\rho \partial^\rho \hat{\eta})_c]\hat{\omega}_c \ \ \alpha \\
&+\tfrac{1}{2}\delta_{ab}g_{\mu\nu}\hat{\eta}_c[\partial^\rho(\hat{D}_\rho \hat{\omega})_c] \ \ (1-\alpha).
\end{aligned} \tag{4.18}$$

Like the fields in the energy-momentum tensor, the fields in the operator $E_{\mu\nu}$ above are bare. The parameter $\alpha$ in the last two lines above is free to vary between 0 and 1 since the matrix elements considered are not sufficient to distinguish between the equations of motion for the ghost and antighost fields.

We find that the following elements of the mixing matrix are compatible with both the two-gluon and two-ghost projections:

$$\begin{aligned}
Z_{GG} =& 1 + \mathcal{O}(g^3) \\
Z_{GA} =& -\frac{1}{2}\frac{1}{\epsilon}\frac{g^2}{16\pi^2}C_A + \mathcal{O}(g^3) \\
Z_{GE} =& \frac{1}{2}\frac{1}{\epsilon}\frac{g^2}{16\pi^2}C_A + \mathcal{O}(g^3) \\
Z_{AG} =& \mathcal{O}(g^3) \\
Z_{AA} =& 1 + \frac{1}{2}\frac{1}{\epsilon}\frac{g^2}{16\pi^2}C_A + \mathcal{O}(g^3) \\
Z_{AE} =& -\frac{1}{2}\frac{1}{\epsilon}\frac{g^2}{16\pi^2}C_A + \mathcal{O}(g^3) \\
Z_{EE} =& 1 + \mathcal{O}(g^2).
\end{aligned} \tag{4.19}$$

The renormalization mixing matrix is triangular to $\mathcal{O}(g^2)$. We do not calculate $Z_{EE}$ beyond the tree level.

## 4.4 BRST Ward Identity (Slavnov-Taylor Identity)

We have seen that the gauge-variant part of the energy-momentum tensor is nonzero in an on-shell matrix element. However, the gauge-variant part of the energy-momentum tensor is BRST-exact, in accordance with general theory and a very simple proof states that physical matrix elements of such operators vanish [13].

In this section, we resolve the contradiction. The proof that physical matrix elements of BRST-exact operators vanish proceeds by using a Ward identity based on the BRST variation to relate the matrix element under study to a particular Green function of the ancestor operator of the BRST-exact operator. This Green function has a manifest factor of zero when put on shell, but the zero is compensated by a quadratic infra-red divergence present only when the matrix element of the ancestor operator is evaluated at zero momentum transfer, as we will now see.

Let us *assume* that the BRST Ward identity holds for unamputated Green functions off shell, calculated at zero momentum transfer.

The BRST variations of the bare fields are given in Eq. (3.7). We need the BRST variation of the renormalized fields in terms of renormalized fields and parameters. This is sometimes referred to as the 'renormalized BRST variation' and is related to the canonical BRST variation by factors of the renormalization constants, Eq. (3.21). The goal in defining a renormalized BRST variation is to obtain UV finite Green functions with renormalized fields. The renormalized constant Grassmann parameter which accomplishes this goal [13] is

$$\delta\xi \equiv Z_A^{-\frac{1}{2}} Z_\eta^{-\frac{1}{2}} \widehat{\delta\xi} \tag{4.20}$$

so we have

$$\begin{aligned}
\delta_{\rm BRST} A_{\mu\ a} &= \left( \frac{Z_\omega^{\frac{1}{2}}}{Z_A^{\frac{1}{2}}} \partial_\mu \omega_a - Z_g Z_\omega^{\frac{1}{2}} g c_{abc} A_{\mu\ b} \omega_c \right) \widehat{\delta\xi} &\Longrightarrow\ & \frac{\delta^r_{\rm BRST}}{\delta\xi} A_{\mu\ a} = Z_0 \partial_\mu \omega_a - Z_g Z_0 Z_A^{\frac{1}{2}} g c_{abc} A_{\mu\ b} \omega_c \\
\delta_{\rm BRST} \omega_a &= -\tfrac{1}{2} Z_g Z_\omega^{\frac{1}{2}} g c_{abc} \omega_b \omega_c \widehat{\delta\xi} &\Longrightarrow\ & \frac{\delta^r_{\rm BRST}}{\delta\xi} \omega_a = -\tfrac{1}{2} Z_g Z_0 Z_A^{\frac{1}{2}} g c_{abc} \omega_b \omega_c \\
\delta_{\rm BRST} \eta_a &= \frac{Z_\lambda Z_A^{\frac{1}{2}}}{Z_\eta^{\frac{1}{2}}} \lambda \partial \cdot A_a \widehat{\delta\xi} &\Longrightarrow\ & \frac{\delta^r_{\rm BRST}}{\delta\xi} \eta_a = \lambda \partial \cdot A_a
\end{aligned} \tag{4.21}$$

The three operators $\frac{\delta^r_{\rm BRST}}{\delta\xi} A_{\mu\ a}$, $\frac{\delta^r_{\rm BRST}}{\delta\xi} \omega_a$, and $\frac{\delta^r_{\rm BRST}}{\delta\xi} \eta_a$ are all finite.

The finite ancestor operator is

$$ancestor\left(\theta_{\mu\nu}^{(GV)}\right) = (\partial_\nu \eta_a) A_{\mu\ a} + (\partial_\mu \eta_a) A_{\nu\ a} - g_{\mu\nu} [\tfrac{1}{2} \eta_a \partial \cdot A_a + (\partial_\rho \eta_a) A_a^\rho] \ , \tag{4.22}$$

where

$$\frac{\delta^r_{\rm BRST}}{\delta\xi} ancestor\left(\theta_{\mu\nu}^{(GV)}\right) = \theta_{\mu\nu}^{(GV)} - \tfrac{1}{2} g_{\mu\nu} \hat{\eta}_a \partial^\rho (\hat{D}_\rho \hat{\omega})_a \ . \tag{4.23}$$

Remember that the fields in the energy-momentum tensor operator are bare. The bare ancestor and the finite (renormalized) ancestor are related by

$$\widehat{ancestor}\left(\theta_{\mu\nu}^{(GV)}\right) = Z_A^{\frac{1}{2}} Z_\eta^{\frac{1}{2}} ancestor\left(\theta_{\mu\nu}^{(GV)}\right) \tag{4.24}$$

Now, the BRST variation of any Green function vanishes. Consider the particular case

$$\delta_{\rm BRST} \langle 0 | T A_{\sigma\ a} ancestor\left(\theta_{\mu\nu}^{(GV)}\right) A_{\tau\ b} | 0 \rangle = 0. \tag{4.25}$$

This gives

$$
\begin{aligned}
0 = \;& \Big\langle 0 \Big| T[\delta_{\mathrm{BRST}} A_{\sigma\ a}] ancestor \left(\theta^{(GV)}_{\mu\nu}\right) A_{\tau\ b} \Big| 0 \Big\rangle \\
&+ \Big\langle 0 \Big| T A_{\sigma\ a} \left[\delta_{\mathrm{BRST}} ancestor \left(\theta^{(GV)}_{\mu\nu}\right)\right] A_{\tau\ b} \Big| 0 \Big\rangle \\
&+ \Big\langle 0 \Big| T A_{\sigma\ a} ancestor \left(\theta^{(GV)}_{\mu\nu}\right) [\delta_{\mathrm{BRST}} A_{\tau\ b}] \Big| 0 \Big\rangle .
\end{aligned}
\tag{4.26}
$$

The vanishing of Eq. (4.25) and the chain rule for the BRST variation can be proven by defining the variation in terms of (anti)commutators with the Noether charge associated with the BRST symmetry [19].

Equation (4.26) becomes

$$
\begin{aligned}
0 = & \Big\langle 0 \Big| T \left[ \left( Z_0 \partial_\sigma \omega_a - Z_g Z_0 Z_A^{\frac{1}{2}} g c_{ade} A_{\sigma\ d} \omega_e \right) \delta\xi \right] ancestor \left(\theta^{(GV)}_{\mu\nu}\right) A_{\tau\ b} \Big| 0 \Big\rangle \\
&+ \Big\langle 0 \Big| T A_{\sigma\ a} \left[ \theta^{(GV)}_{\mu\nu} \delta\xi \right] A_{\tau\ b} \Big| 0 \Big\rangle \\
&+ \Big\langle 0 \Big| T A_{\sigma\ a} \left[ -\tfrac{1}{2} g_{\mu\nu} \hat{\eta}_c \partial^\rho (\hat{D}_\rho \hat{\omega})_c \delta\xi \right] A_{\tau\ b} \Big| 0 \Big\rangle \\
&+ \Big\langle 0 \Big| T A_{\sigma\ a} ancestor \left(\theta^{(GV)}_{\mu\nu}\right) \left[ (Z_0 \partial_\tau \omega_b - Z_g Z_0 Z_A^{\frac{1}{2}} g c_{bde} A_{\tau\ d} \omega_e) \delta\xi \right] \Big| 0 \Big\rangle .
\end{aligned}
\tag{4.27}
$$

The constant Grassmann parameter, $\delta\xi$, can be factored out of Eq. (4.27) above if it is anti-commuted through the ancestor operator which has a Grassmann parity of 1 (because each term contains one antighost field). This is responsible for the relative minus sign below.

$$
\begin{aligned}
0 = (-1) & \Big\langle 0 \Big| T \left[ Z_0 \partial_\sigma \omega_a - Z_g Z_0 Z_A^{\frac{1}{2}} g c_{ade} A_{\sigma\ d} \omega_e \right] ancestor \left(\theta^{(GV)}_{\mu\nu}\right) A_{\tau\ b} \Big| 0 \Big\rangle \\
&+ \Big\langle 0 \Big| T A_{\sigma\ a} \theta^{(GV)}_{\mu\nu} A_{\tau\ b} \Big| 0 \Big\rangle \\
&+ \Big\langle 0 \Big| T A_{\sigma\ a} \left[ -\tfrac{1}{2} g_{\mu\nu} \hat{\eta}_c \partial^\rho (\hat{D}_\rho \hat{\omega})_c \right] A_{\tau\ b} \Big| 0 \Big\rangle \\
&+ \Big\langle 0 \Big| T A_{\sigma\ a} ancestor \left(\theta^{(GV)}_{\mu\nu}\right) \left[ Z_0 \partial_\tau \omega_b - Z_g Z_0 Z_A^{\frac{1}{2}} g c_{bde} A_{\tau\ d} \omega_e \right] \Big| 0 \Big\rangle .
\end{aligned}
\tag{4.28}
$$

One must also keep in mind that the order of the Grassmann fields, $\hat{\eta}$ (in the ancestor operator) and $\hat{\omega}$, in the last two lines above is opposite to the canonical ordering and that

$$\langle 0|T\hat{\eta}\hat{\omega}(Operator)|0\rangle = -\langle 0|T\hat{\omega}\hat{\eta}(Operator)|0\rangle \tag{4.29}$$

It is now obvious that there exists an alternate calculation which will provide the two-gluon physical matrix element of $\theta^{(GV)}_{\mu\nu}$.

$$\begin{aligned}\left\langle 0\left|TA_{\sigma\ a}\theta^{(GV)}_{\mu\nu}A_{\tau\ b}\right|0\right\rangle =& \left\langle 0\left|T\left[Z_0\partial_\sigma\omega_a - Z_gZ_0Z_A^{\frac{1}{2}}gc_{ade}A_{\sigma\ d}\omega_e\right] ancestor\left(\theta^{(GV)}_{\mu\nu}\right)A_{\tau\ b}\right|0\right\rangle \\ &- \left\langle 0\left|TA_{\sigma\ a}\left[-\tfrac{1}{2}g_{\mu\nu}\hat{\eta}_c\partial^\rho(\hat{D}_\rho\hat{\omega})_c\right]A_{\tau\ b}\right|0\right\rangle \\ &- \left\langle 0\left|TA_{\sigma\ a} ancestor\left(\theta^{(GV)}_{\mu\nu}\right)\left[Z_0\partial_\tau\omega_b - Z_gZ_0Z_A^{\frac{1}{2}}gc_{bde}A_{\tau\ d}\omega_e\right]\right|0\right\rangle\end{aligned} \tag{4.30}$$

The usual proof that this matrix element vanishes on shell relies on the assumption that the triangle graph depicted on the bottom of page 46 in the Appendix does not contain a quadratic infra-red singularity. This graph and another like it contain an unusual vertex, one gluon and one ghost at the same space-time point. There is no external line, therefore nothing to amputate, but the amputation procedure for other graphs, which have ghost legs, requires that all diagrams be multiplied by the inverse of a ghost propagator, which of course is proportional to $p^2$. If the diagrams contain at worst logarithmic IR divergences, then the additional factor of $p^2$ will cause them to vanish on shell. The diagrams with an external ghost line have a derivative acting on the ghost field. In momentum space, the derivative becomes a factor of $p^\sigma$ which gives zero when contracted with the physical polarization vector, $\epsilon_\sigma$, associated with the gluon leg.

Explicit calculation of the relevant graph, with the result Eq. (A.9), shows that there is, in fact, a $\frac{1}{p^2}$ divergence which cancels against the inverse of the ghost propagator introduced in the amputation procedure, as we claimed at the beginning of this section. This IR pole occurs only at zero momentum transfer.

## 5 Conclusion

We have seen by explicit calculation that one of the central results of Joglekar and Lee, that physical matrix elements of BRST-exact operators must vanish, fails at one loop order. We give a form for the BRST ancestors of the alien operators required in the renormalization of the covariant gluon operator. Our alien operators are then manifestly BRST-exact whereas the basis of alien operators proposed by Dixon and Taylor (those used by Hamberg and van Neerven) are not BRST-exact. The Dixon and Taylor set of alien operators are not guaranteed to vanish in physical matrix elements.

We have verified the predictions of Freedman *et al.* on the finiteness of the energy-momentum tensor in both gluon and ghost two-point functions to one loop order at zero momentum transfer by evaluating diagrams with a BRST-exact alien operator insertion.

The BRST Ward identity demonstrates where the proof of the Joglekar–Lee theorem breaks down. Taking the momentum transfer to zero too soon introduces spurious infra-red divergences which cancel factors of zero on which the proof relies. Calculations performed using the Dixon and Taylor set of alien operators cannot by analyzed through the BRST Ward identity.

The physical region of interest in almost all calculations involving the renormalization of composite operators, such as the calculations required in the operator product expansion, is the exceptional point of zero momentum transfer. To expedite the computation, one sets the momentum transfer to zero at the very beginning, thereby eliminating one scale from the problem. In some calculations involving final state cuts, it is not clear how one would generalize to nonzero momentum transfer.

The alternative is to keep the momentum transfer arbitrary until after the Feynman graphs have been evaluated, and only then to set the momentum transfer to zero. With this procedure, the Joglekar–Lee theorem should apply, making it unnecessary to compute the graphs containing the alien operator insertion. The price to be paid, of course, is the introduction of another momentum scale into the problem and a corresponding increase in the complexity and volume of the analysis. However, it is not obvious that the limit of zero momentum transfer is the only source of contradiction with theory. A recent calculation by Harris and Smith [16] at nonzero momentum transfer indicates that the discrepancy persists.

There has been a sense of disquiet in the literature about zero momentum transfer for a long time. Joglekar mentions in the concluding section of [24] that, at the exceptional momentum point, $Q = 0$, matrix elements of gauge-invariant operators lose some of the properties that make them manageable at nonzero momentum transfer. C. Lee [25] works with the twist-2 piece of the energy-momentum tensor to show that certain pieces of the calculation at zero momentum transfer can yield useful information, that is the coefficients of certain terms are the same, independent of the momentum transfer. He calculates only the pole terms of the two-gluon Green function at one-loop order at both zero and nonzero momentum transfer. The unease was certainly justified; some results hold while others fail utterly. It is not unreasonable to question all calculations performed when the limit of zero momentum transfer was applied initially, and such calculations are the mainstay of perturbative QCD.

## Acknowledgments

We would like to thank Jack Smith and Brian Harris of SUNY Stonybrook and Willie van Neerven of the University of Leiden, the Netherlands, for enlightening discussions. In addition, thanks are due Laura Weinkauf of Penn State for technical help with the combinatorics of the diagrams and the FORM code.

## Appendix

### A.1 Right Derivatives

Right derivatives [26] are such that

$$\frac{\partial^r (XY)}{\partial Z} = X \frac{\partial^r Y}{\partial Z} + (-1)^{P_Y P_Z} \left( \frac{\partial^r X}{\partial Z} \right) Y, \tag{A.1}$$

where $P_Y$ is the 'Grassmann parity' of the quantity $Y$. The (anti)ghost field components have Grassmann parity 1, while the c-number parameters and boson field components have Grassmann parity 0. Fermion field components would be assigned Grassmann parity 1.

## A.2 Complete Off-Shell Calculations for Two-Gluon Amputated Green Functions

In this part of the Appendix, we give the full Lorentz structure for the amputated Green functions of the energy-momentum tensor operators with two gluon fields off mass-shell, at zero momentum transfer, to one loop order. The $\frac{1}{\epsilon}$ poles are purely UV divergences while all of the IR divergences (as $p^2 \to 0$) are seen as logarithms.

### A.2.1 Entire Energy-Momentum Tensor

$$
\begin{aligned}
\langle 0|TA_{\sigma\ a}\theta_{\mu\nu}A_{\tau\ b}|0\rangle_{Amputated} = \delta_{ab}[p^2(g_{\sigma\tau}g_{\mu\nu} - g_{\sigma\mu}g_{\tau\nu} - g_{\sigma\nu}g_{\tau\mu})\\
-2p_\mu p_\nu g_{\sigma\tau} + p_\tau p_\nu g_{\sigma\mu} + p_\tau p_\mu g_{\sigma\nu} + p_\sigma p_\nu g_{\tau\mu} + p_\sigma p_\mu g_{\tau\nu} - p_\sigma p_\tau g_{\mu\nu}\\
-\lambda(p_\tau p_\nu g_{\sigma\mu} + p_\tau p_\mu g_{\sigma\nu} + p_\sigma p_\nu g_{\tau\mu} + p_\sigma p_\mu g_{\tau\nu} - p_\sigma p_\tau g_{\mu\nu})]\\
+\frac{g^2}{16\pi^2}C_A\delta_{ab}\left\{
\begin{array}{l}
-\frac{1}{4}\left(1-\frac{1}{\lambda}\right)^2
\left[\begin{array}{l}
p^2(g_{\sigma\tau}g_{\mu\nu} - g_{\sigma\mu}g_{\tau\nu} - g_{\sigma\nu}g_{\tau\mu})\\
-2p_\mu p_\nu g_{\sigma\tau} + p_\tau p_\nu g_{\sigma\mu} + p_\tau p_\mu g_{\sigma\nu}\\
+p_\sigma p_\nu g_{\tau\mu} + p_\sigma p_\mu g_{\tau\nu} - p_\sigma p_\tau g_{\mu\nu}
\end{array}\right]\\
+\left(1-\frac{1}{\lambda}\right)
\left[\begin{array}{l}
\frac{p_\sigma p_\tau p_\mu p_\nu}{p^2}\\
-\left[3+\ln\left(\frac{-p^2}{\bar{\mu}^2}\right)\right]p_\mu p_\nu g_{\sigma\tau}\\
+\left[1+\frac{1}{2}\ln\left(\frac{-p^2}{\bar{\mu}^2}\right)\right]
\left[\begin{array}{l}
p^2(g_{\sigma\tau}g_{\mu\nu} - g_{\sigma\mu}g_{\tau\nu} - g_{\sigma\nu}g_{\tau\mu})\\
+p_\tau p_\nu g_{\sigma\mu} + p_\tau p_\mu g_{\sigma\nu} + p_\sigma p_\nu g_{\tau\mu}\\
+p_\sigma p_\mu g_{\tau\nu} - p_\sigma p_\tau g_{\mu\nu}
\end{array}\right]
\end{array}\right]\\
+\frac{10}{3}\frac{p_\sigma p_\tau p_\mu p_\nu}{p^2}\\
+\left[\frac{32}{9}-\frac{10}{3}\ln\left(\frac{-p^2}{\bar{\mu}^2}\right)\right]p_\mu p_\nu g_{\sigma\tau}\\
+\left[-\frac{31}{9}+\frac{5}{3}\ln\left(\frac{-p^2}{\bar{\mu}^2}\right)\right]
\left[\begin{array}{l}
p^2(g_{\sigma\tau}g_{\mu\nu} - g_{\sigma\mu}g_{\tau\nu} - g_{\sigma\nu}g_{\tau\mu})\\
+p_\tau p_\nu g_{\sigma\mu} + p_\tau p_\mu g_{\sigma\nu} + p_\sigma p_\nu g_{\tau\mu}\\
+p_\sigma p_\mu g_{\tau\nu} - p_\sigma p_\tau g_{\mu\nu}
\end{array}\right]
\end{array}
\right\}\\
+\mathcal{O}(g^4)
\end{aligned}
\tag{A.2}
$$

To make the comparison with Hamberg and van Neerven more transparent, we present the twist-2 (spin-2) piece of this amputated Green function in which the free Lorentz indices of the inserted operator, $\mu$ and $\nu$, are contracted with a null-vector, $\Delta$:

$$
\begin{aligned}
\langle 0|TA_{\sigma\ a}\Delta^\mu\theta_{\mu\nu}\Delta^\nu A_{\tau\ b}|0\rangle_{Amputated} &= 2\delta_{ab}[-p^2\Delta_\sigma\Delta_\tau-(p\cdot\Delta)^2g_{\sigma\tau}+(1-\lambda)(p\cdot\Delta)(p_\sigma\Delta_\tau+p_\tau\Delta_\sigma)] \\
&+\tfrac{g^2}{16\pi^2}C_A\delta_{ab}\left\{\begin{array}{l}
-\frac{1}{2}\left(1-\frac{1}{\lambda}\right)^2[-p^2\Delta_\sigma\Delta_\tau-(p\cdot\Delta)^2g_{\sigma\tau}+(p\cdot\Delta)(p_\sigma\Delta_\tau+p_\tau\Delta_\sigma)] \\
+\left(1-\frac{1}{\lambda}\right)\left[\begin{array}{l}
\dfrac{p_\sigma p_\tau(p\cdot\Delta)^2}{p^2} \\
-\left[3+\ln\left(\dfrac{-p^2}{\bar\mu^2}\right)\right](p\cdot\Delta)^2g_{\sigma\tau} \\
+\left[2+\ln\left(\dfrac{-p^2}{\bar\mu^2}\right)\right]\left[\begin{array}{l}-p^2\Delta_\sigma\Delta_\tau \\ +(p\cdot\Delta)(p_\sigma\Delta_\tau+p_\tau\Delta_\sigma)\end{array}\right]
\end{array}\right] \\
+\frac{10}{3}\dfrac{p_\sigma p_\tau(p\cdot\Delta)^2}{p^2}+\left[\frac{32}{9}-\frac{10}{3}\ln\left(\dfrac{-p^2}{\bar\mu^2}\right)\right](p\cdot\Delta)^2g_{\sigma\tau} \\
+\left[-\frac{62}{9}+\frac{10}{3}\ln\left(\dfrac{-p^2}{\bar\mu^2}\right)\right][-p^2\Delta_\sigma\Delta_\tau+(p\cdot\Delta)(p_\sigma\Delta_\tau+p_\tau\Delta_\sigma)]
\end{array}\right\} \\
&+\mathcal{O}(g^4)
\end{aligned}
\tag{A.3}
$$

### A.2.2 Gauge-Invariant Part

$$
\begin{aligned}
\langle 0|TA_{\sigma\ a}\theta^{(GI)}_{\mu\nu}A_{\tau\ b}|0\rangle_{Amputated} = {} & \tfrac{1}{\epsilon}\tfrac{g^2}{16\pi^2}C_A\delta_{ab}[p^2(-\tfrac{1}{2}g_{\sigma\tau}g_{\mu\nu}+g_{\sigma\mu}g_{\tau\nu}+g_{\sigma\nu}g_{\tau\mu}) \\
& -\tfrac{1}{2}(p_\tau p_\nu g_{\sigma\mu}+p_\tau p_\mu g_{\sigma\nu}+p_\sigma p_\nu g_{\tau\mu}+p_\sigma p_\mu g_{\tau\nu}-p_\sigma p_\tau g_{\mu\nu})] \\
& +\delta_{ab}[p^2(g_{\sigma\tau}g_{\mu\nu}-g_{\sigma\mu}g_{\tau\nu}-g_{\sigma\nu}g_{\tau\mu}) \\
& -2p_\mu p_\nu g_{\sigma\tau}+p_\tau p_\nu g_{\sigma\mu}+p_\tau p_\mu g_{\sigma\nu}+p_\sigma p_\nu g_{\tau\mu}+p_\sigma p_\mu g_{\tau\nu}-p_\sigma p_\tau g_{\mu\nu}] \\
& +\tfrac{g^2}{16\pi^2}C_A\delta_{ab}\left\{\begin{array}{l}
-\tfrac{1}{2}\left(1-\tfrac{1}{\lambda}\right)^2\left[\begin{array}{l} p^2(g_{\sigma\tau}g_{\mu\nu}-g_{\sigma\mu}g_{\tau\nu}-g_{\sigma\nu}g_{\tau\mu}) \\ -2p_\mu p_\nu g_{\sigma\tau}+p_\tau p_\nu g_{\sigma\mu}+p_\tau p_\mu g_{\sigma\nu} \\ +p_\sigma p_\nu g_{\tau\mu}+p_\sigma p_\mu g_{\tau\nu}-p_\sigma p_\tau g_{\mu\nu}\end{array}\right] \\
+\left(1-\tfrac{1}{\lambda}\right)\left[\begin{array}{l} 2\dfrac{p_\sigma p_\tau p_\mu p_\nu}{p^2}-\left[6+\ln\left(\dfrac{-p^2}{\bar{\mu}^2}\right)\right]p_\mu p_\nu g_{\sigma\tau} \\ +\left[\tfrac{5}{2}+\tfrac{1}{2}\ln\left(\dfrac{-p^2}{\bar{\mu}^2}\right)\right](p^2 g_{\sigma\tau}g_{\mu\nu}-p_\sigma p_\tau g_{\mu\nu}) \\ +\left[2+\tfrac{1}{2}\ln\left(\dfrac{-p^2}{\bar{\mu}^2}\right)\right]\left[\begin{array}{l} -p^2(g_{\sigma\mu}g_{\tau\nu}+g_{\sigma\nu}g_{\tau\mu}) \\ +p_\tau p_\nu g_{\sigma\mu}+p_\tau p_\mu g_{\sigma\nu} \\ +p_\sigma p_\nu g_{\tau\mu}+p_\sigma p_\mu g_{\tau\nu}\end{array}\right]\end{array}\right] \\
-\tfrac{2}{3}\dfrac{p_\sigma p_\tau p_\mu p_\nu}{p^2}+\left[\tfrac{86}{9}-\tfrac{10}{3}\ln\left(\dfrac{-p^2}{\bar{\mu}^2}\right)\right]p_\mu p_\nu g_{\sigma\tau} \\
+\left[-\tfrac{58}{9}+\tfrac{13}{6}\ln\left(\dfrac{-p^2}{\bar{\mu}^2}\right)\right](p^2 g_{\sigma\tau}g_{\mu\nu}-p_\sigma p_\tau g_{\mu\nu}) \\
+\left[\tfrac{49}{9}-\tfrac{8}{3}\ln\left(\dfrac{-p^2}{\bar{\mu}^2}\right)\right]p^2(g_{\sigma\mu}g_{\tau\nu}+g_{\sigma\nu}g_{\tau\mu}) \\
+\left[-\tfrac{89}{18}+\tfrac{13}{6}\ln\left(\dfrac{-p^2}{\bar{\mu}^2}\right)\right](p_\tau p_\nu g_{\sigma\mu}+p_\tau p_\mu g_{\sigma\nu}+p_\sigma p_\nu g_{\tau\mu}+p_\sigma p_\mu g_{\tau\nu})
\end{array}\right\} \\
& +\mathcal{O}(g^4)
\end{aligned}
\tag{A.4}
$$

The twist-2 (spin-2) piece of this amputated Green function in which the free Lorentz indices of the inserted operator, $\mu$ and $\nu$, are contracted with a null-vector, $\Delta$, is

$$
\begin{aligned}
\langle 0|T A_{\sigma\ a}\Delta^\mu \theta^{(GI)}_{\mu\nu}\Delta^\nu A_{\tau\ b}|0\rangle_{Amputated} &= \tfrac{1}{\epsilon}\tfrac{g^2}{16\pi^2} C_A \delta_{ab}[2p^2\Delta_\sigma\Delta_\tau - (p\cdot\Delta)(p_\sigma\Delta_\tau + p_\tau\Delta_\sigma)] \\
&+2\delta_{ab}[-p^2\Delta_\sigma\Delta_\tau - (p\cdot\Delta)^2 g_{\sigma\tau} + (p\cdot\Delta)(p_\sigma\Delta_\tau + p_\tau\Delta_\sigma)] \\
&+\tfrac{g^2}{16\pi^2}C_A\delta_{ab}\left\{
\begin{array}{l}
-\left(1-\frac{1}{\lambda}\right)^2\left[-p^2\Delta_\sigma\Delta_\tau - (p\cdot\Delta)^2 g_{\sigma\tau} + (p\cdot\Delta)(p_\sigma\Delta_\tau + p_\tau\Delta_\sigma)\right] \\
+\left(1-\frac{1}{\lambda}\right)\left[
\begin{array}{l}
2\dfrac{p_\sigma p_\tau (p\cdot\Delta)^2}{p^2} - \left[6+\ln\left(\dfrac{-p^2}{\bar\mu^2}\right)\right](p\cdot\Delta)^2 g_{\sigma\tau} \\
+\left[4+\ln\left(\dfrac{-p^2}{\bar\mu^2}\right)\right]\left[-p^2\Delta_\sigma\Delta_\tau + (p\cdot\Delta)(p_\sigma\Delta_\tau + p_\tau\Delta_\sigma)\right]
\end{array}\right] \\
-\frac{2}{3}\dfrac{p_\sigma p_\tau (p\cdot\Delta)^2}{p^2} + \left[\frac{86}{9} - \frac{10}{3}\ln\left(\dfrac{-p^2}{\bar\mu^2}\right)\right](p\cdot\Delta)^2 g_{\sigma\tau} \\
+\left[\frac{98}{9} - \frac{16}{3}\ln\left(\dfrac{-p^2}{\bar\mu^2}\right)\right]p^2\Delta_\sigma\Delta_\tau \\
+\left[-\frac{89}{9} + \frac{13}{3}\ln\left(\dfrac{-p^2}{\bar\mu^2}\right)\right](p\cdot\Delta)(p_\sigma\Delta_\tau + p_\tau\Delta_\sigma)
\end{array}\right\} \\
&+\mathcal{O}(g^4)
\end{aligned}
\tag{A.5}
$$

### A.2.3 Gauge-Variant (Alien) Part

$$
\begin{aligned}
\langle 0|TA_{\sigma\ a}\theta^{(GV)}_{\mu\nu}A_{\tau\ b}|0\rangle_{Amputated} = &-\frac{1}{\epsilon}\frac{g^2}{16\pi^2}C_A\delta_{ab}[p^2(-\tfrac{1}{2}g_{\sigma\tau}g_{\mu\nu}+g_{\sigma\mu}g_{\tau\nu}+g_{\sigma\nu}g_{\tau\mu}) \\
&-\tfrac{1}{2}(p_\tau p_\nu g_{\sigma\mu}+p_\tau p_\mu g_{\sigma\nu}+p_\sigma p_\nu g_{\tau\mu}+p_\sigma p_\mu g_{\tau\nu}-p_\sigma p_\tau g_{\mu\nu})] \\
&-\lambda\delta_{ab}(p_\tau p_\nu g_{\sigma\mu}+p_\tau p_\mu g_{\sigma\nu}+p_\sigma p_\nu g_{\tau\mu}+p_\sigma p_\mu g_{\tau\nu}-p_\sigma p_\tau g_{\mu\nu}) \\
&+\frac{g^2}{16\pi^2}C_A\delta_{ab}\left\{\begin{array}{l}
\frac{1}{4}\left(1-\frac{1}{\lambda}\right)^2\left[\begin{array}{l} p^2(g_{\sigma\tau}g_{\mu\nu}-g_{\sigma\mu}g_{\tau\nu}-g_{\sigma\nu}g_{\tau\mu}) \\ -2p_\mu p_\nu g_{\sigma\tau}+p_\tau p_\nu g_{\sigma\mu}+p_\tau p_\mu g_{\sigma\nu} \\ +p_\sigma p_\nu g_{\tau\mu}+p_\sigma p_\mu g_{\tau\nu}-p_\sigma p_\tau g_{\mu\nu}\end{array}\right] \\
+\left(1-\frac{1}{\lambda}\right)\left[\begin{array}{l} -\dfrac{p_\sigma p_\tau p_\mu p_\nu}{p^2} \\ +p^2(-\frac{3}{2}g_{\sigma\tau}g_{\mu\nu}+g_{\sigma\mu}g_{\tau\nu}+g_{\sigma\nu}g_{\tau\mu}) \\ +3p_\mu p_\nu g_{\sigma\tau}-p_\tau p_\nu g_{\sigma\mu}-p_\tau p_\mu g_{\sigma\nu} \\ -p_\sigma p_\nu g_{\tau\mu}-p_\sigma p_\mu g_{\tau\nu}+\frac{3}{2}p_\sigma p_\tau g_{\mu\nu}\end{array}\right] \\
+4\dfrac{p_\sigma p_\tau p_\mu p_\nu}{p^2}-6p_\mu p_\nu g_{\sigma\tau} \\
+\left[3-\frac{1}{2}\ln\left(\dfrac{-p^2}{\bar{\mu}^2}\right)\right](p^2 g_{\sigma\tau}g_{\mu\nu}-p_\sigma p_\tau g_{\mu\nu}) \\
+\left[-2+\ln\left(\dfrac{-p^2}{\bar{\mu}^2}\right)\right]p^2(g_{\sigma\mu}g_{\tau\nu}+g_{\sigma\nu}g_{\tau\mu}) \\
+\left[\frac{3}{2}-\frac{1}{2}\ln\left(\dfrac{-p^2}{\bar{\mu}^2}\right)\right](p_\tau p_\nu g_{\sigma\mu}+p_\tau p_\mu g_{\sigma\nu}+p_\sigma p_\nu g_{\tau\mu}+p_\sigma p_\mu g_{\tau\nu})
\end{array}\right\} \\
&+\mathcal{O}(g^4)
\end{aligned}
\tag{A.6}
$$

The twist-2 (spin-2) piece of this amputated Green function in which the free Lorentz indices of the inserted operator, $\mu$ and $\nu$, are contracted with a null-vector, $\Delta$, is

$$
\begin{aligned}
\langle 0|TA_{\sigma\ a}\Delta^\mu\theta^{(GV)}_{\mu\nu}\Delta^\nu A_{\tau\ b}|0\rangle_{Amputated} = {} & -\tfrac{1}{\epsilon}\tfrac{g^2}{16\pi^2}C_A\delta_{ab}[2p^2\Delta_\sigma\Delta_\tau - (p\cdot\Delta)(p_\sigma\Delta_\tau + p_\tau\Delta_\sigma)] \\
& -2\lambda\delta_{ab}(p\cdot\Delta)(p_\sigma\Delta_\tau + p_\tau\Delta_\sigma) \\
& +\tfrac{g^2}{16\pi^2}C_A\delta_{ab}\left\{\begin{array}{l}
\frac{1}{2}\left(1-\frac{1}{\lambda}\right)^2\left[-p^2\Delta_\sigma\Delta_\tau - (p\cdot\Delta)^2 g_{\sigma\tau} + (p\cdot\Delta)(p_\sigma\Delta_\tau + p_\tau\Delta_\sigma)\right] \\
+\left(1-\frac{1}{\lambda}\right)\left[\begin{array}{l} -\dfrac{p_\sigma p_\tau(p\cdot\Delta)^2}{p^2} \\ +2p^2\Delta_\sigma\Delta_\tau + 3(p\cdot\Delta)^2 g_{\sigma\tau} \\ -2(p\cdot\Delta)(p_\sigma\Delta_\tau + p_\tau\Delta_\sigma) \end{array}\right] \\
+4\dfrac{p_\sigma p_\tau(p\cdot\Delta)^2}{p^2} - 6(p\cdot\Delta)^2 g_{\sigma\tau} + \left[-4 + 2\ln\left(\dfrac{-p^2}{\bar{\mu}^2}\right)\right]p^2\Delta_\sigma\Delta_\tau \\
+\left[3 - \ln\left(\dfrac{-p^2}{\bar{\mu}^2}\right)\right](p\cdot\Delta)(p_\sigma\Delta_\tau + p_\tau\Delta_\sigma)
\end{array}\right\} \\
& +\mathcal{O}(g^4)
\end{aligned}
\tag{A.7}
$$

## A.3 Unamputated Two-Gluon Green Function of Alien Piece of the Energy-Momentum Tensor Off-Shell

In order to use the BRST Ward identity (which is valid off mass-shell), we need the unamputated Green function which we obtain from the amputated Green function, Eq. (A.6), by attaching the dressed external gluon propagators, Eq. (3.24a).[13]

$$
\begin{aligned}
&\langle 0|TA_{\sigma\ a}\theta^{(GV)}_{\mu\nu}A_{\tau\ b}|0\rangle = \mathcal{D}^{\sigma\tau'}_{ab'}(p)\langle 0|TA_{\sigma'\ a'}\theta^{(GV)}_{\mu\nu}A_{\tau'\ b'}|0\rangle_{Amputated}\mathcal{D}^{\sigma'\tau}_{a'b}(p)\\
&=\frac{1}{\epsilon}\frac{g^2}{16\pi^2}C_A\delta_{ab}\frac{1}{(p^2)^2}\left\{\begin{array}{l}
\left(1-\frac{1}{\lambda}\right)\left[2\frac{p_\sigma p_\tau p_\mu p_\nu}{p^2}-\frac{1}{2}(p_\tau p_\nu g_{\sigma\mu}+p_\tau p_\mu g_{\sigma\nu}+p_\sigma p_\nu g_{\tau\mu}+p_\sigma p_\mu g_{\tau\nu})\right]\\
+p^2(-\frac{1}{2}g_{\sigma\tau}g_{\mu\nu}+g_{\sigma\mu}g_{\tau\nu}+g_{\sigma\nu}g_{\tau\mu})\\
-\frac{1}{2}(p_\tau p_\nu g_{\sigma\mu}+p_\tau p_\mu g_{\sigma\nu}+p_\sigma p_\nu g_{\tau\mu}+p_\sigma p_\mu g_{\tau\nu}-p_\sigma p_\tau g_{\mu\nu})
\end{array}\right\}\\
&\quad+\delta_{ab}\frac{1}{(p^2)^2}\left[\begin{array}{l}
-4\left(1-\frac{1}{\lambda}\right)\frac{p_\sigma p_\tau p_\mu p_\nu}{p^2}\\
+p_\tau p_\nu g_{\sigma\mu}+p_\tau p_\mu g_{\sigma\nu}+p_\sigma p_\nu g_{\tau\mu}+p_\sigma p_\mu g_{\tau\nu}-p_\sigma p_\tau g_{\mu\nu}
\end{array}\right]\\
&\quad+\frac{g^2}{16\pi^2}C_A\delta_{ab}\frac{1}{(p^2)^2}\left\{\begin{array}{l}
\left(1-\frac{1}{\lambda}\right)^2\left[\begin{array}{l}-\frac{p_\sigma p_\tau p_\mu p_\nu}{p^2}+\frac{1}{4}p^2(-g_{\sigma\tau}g_{\mu\nu}+g_{\sigma\mu}g_{\tau\nu}+g_{\sigma\nu}g_{\tau\mu})\\ +\frac{1}{2}p_\mu p_\nu g_{\sigma\tau}+\frac{1}{4}p_\sigma p_\tau g_{\mu\nu}\end{array}\right]\\
+\left(1-\frac{1}{\lambda}\right)\left[\begin{array}{l}7\frac{p_\sigma p_\tau p_\mu p_\nu}{p^2}-p^2(-\frac{3}{2}g_{\sigma\tau}g_{\mu\nu}+g_{\sigma\mu}g_{\tau\nu}+g_{\sigma\nu}g_{\tau\mu})\\ -\frac{1}{2}(p_\tau p_\nu g_{\sigma\mu}+p_\tau p_\mu g_{\sigma\nu}+p_\sigma p_\nu g_{\tau\mu}+p_\sigma p_\mu g_{\tau\nu})\\ -3p_\mu p_\nu g_{\sigma\tau}-\frac{3}{2}p_\sigma p_\tau g_{\mu\nu}\end{array}\right]\\
+\left[-\frac{160}{9}+\frac{20}{3}\ln\left(\frac{-p^2}{\bar{\mu}^2}\right)\right]\frac{p_\sigma p_\tau p_\mu p_\nu}{p^2}+\left[-3+\frac{1}{2}\ln\left(\frac{-p^2}{\bar{\mu}^2}\right)\right](p^2 g_{\sigma\tau}g_{\mu\nu}-p_\sigma p_\tau g_{\mu\nu})\\
+\left[2-\ln\left(\frac{-p^2}{\bar{\mu}^2}\right)\right]p^2(g_{\sigma\mu}g_{\tau\nu}+g_{\sigma\nu}g_{\tau\mu})+6p_\mu p_\nu g_{\sigma\tau}\\
+\left[\frac{35}{18}-\frac{7}{6}\ln\left(\frac{-p^2}{\bar{\mu}^2}\right)\right](p_\tau p_\nu g_{\sigma\mu}+p_\tau p_\mu g_{\sigma\nu}+p_\sigma p_\nu g_{\tau\mu}+p_\sigma p_\mu g_{\tau\nu})
\end{array}\right\}\\
&\quad+\mathcal{O}(g^4)
\end{aligned}
\tag{A.8}
$$

[13]The modified LSZ prescription, rather than the dressed external propagators, would have been used to go on shell.

## A.4 Feynman Diagrams

The following graphs contribute to the matrix elements at one-loop order. The external propagators are amputated; one-loop corrections to the legs are handled by the modified LSZ reduction procedure explained in Section 3.6 to go on shell or by attaching dressed external propagators as in Section A.3 to remain off shell. The inserted composite operator is represented by the symbol $\otimes$. In this section, $\otimes = \theta_{\mu\nu}$, $\theta_{\mu\nu}^{(GI)}$, or $\theta_{\mu\nu}^{(GV)}$.

### A.4.1 Gluon Two-point Function

The Born graph:

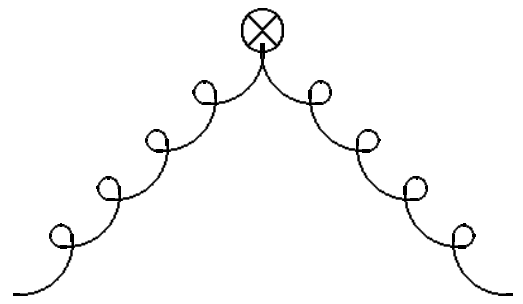

The order $g^2$ graphs:

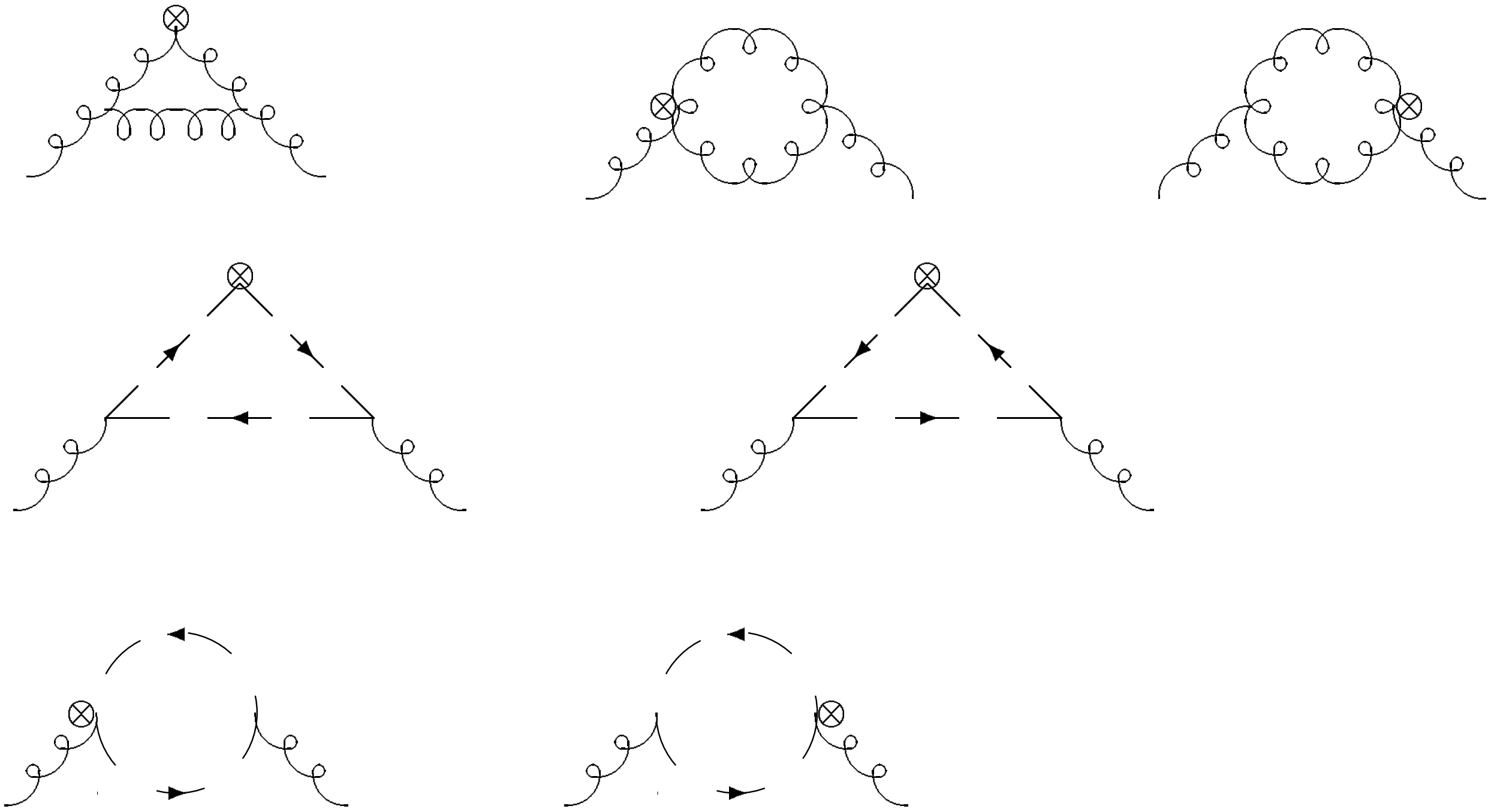

The following diagram vanishes for zero momentum transfer, $Q$, but contributes in the nonzero momentum transfer case:

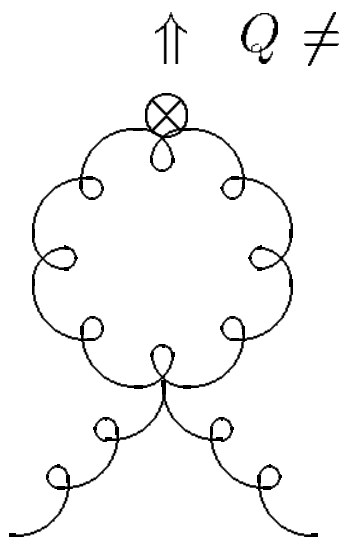

The following diagram always vanishes in dimensional regularization, regardless of the momentum transfer, because the integral contains no momentum scale:

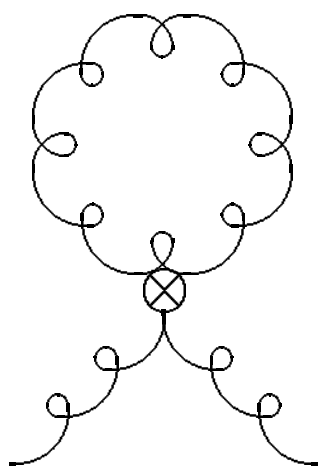

$$\equiv 0 \quad \forall Q$$

### A.4.2 Ghost Two-point Function

The Born graph:

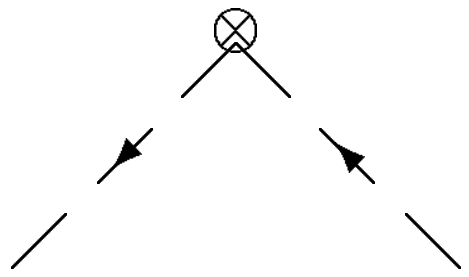

The order $g^2$ graphs:

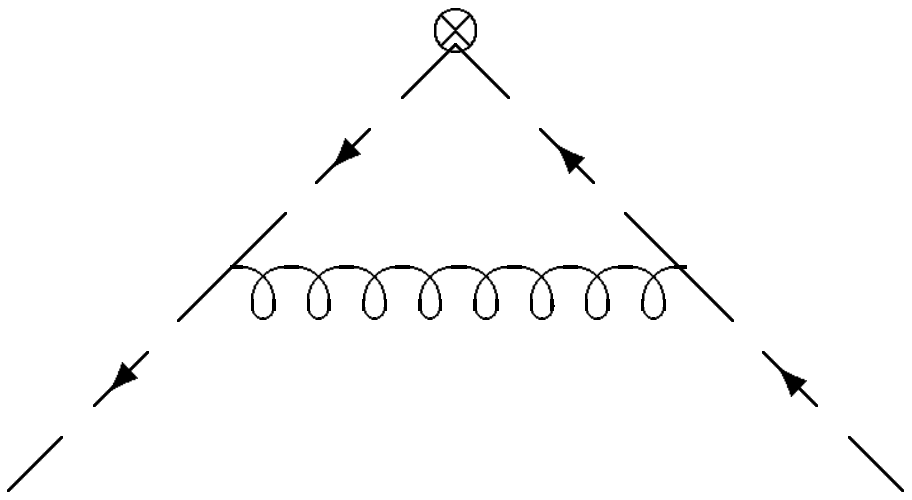

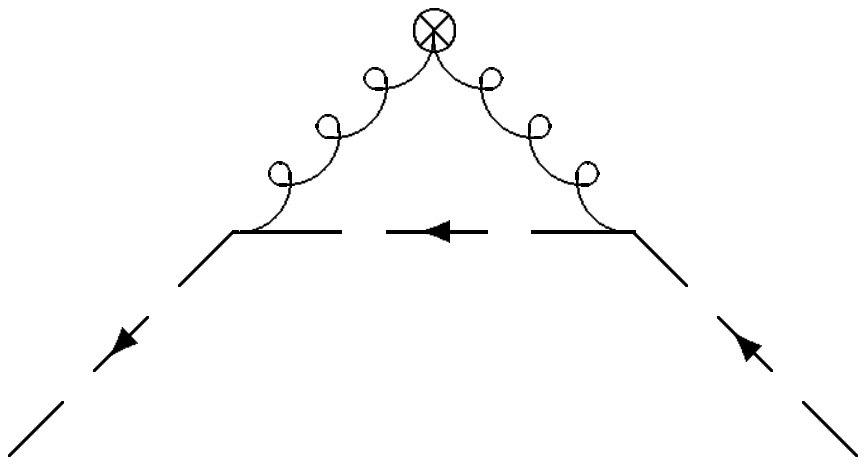

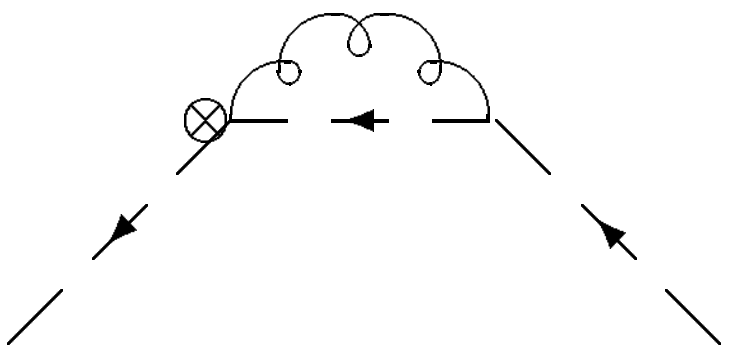

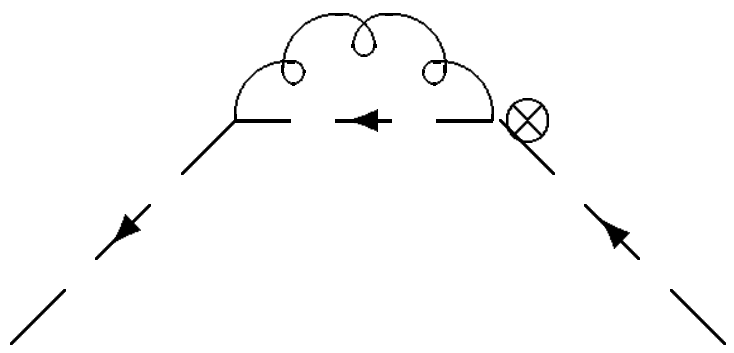

### A.4.3 The BRST Ward Identity Graphs

In this section, the symbol $\otimes$ stands for the renormalized BRST ancestor of $\theta_{\mu\nu}^{(GV)}$, (Eq. 4.22).

The Born graph:

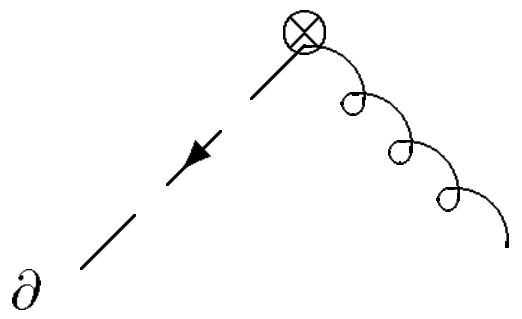

The order $g^2$ graphs:

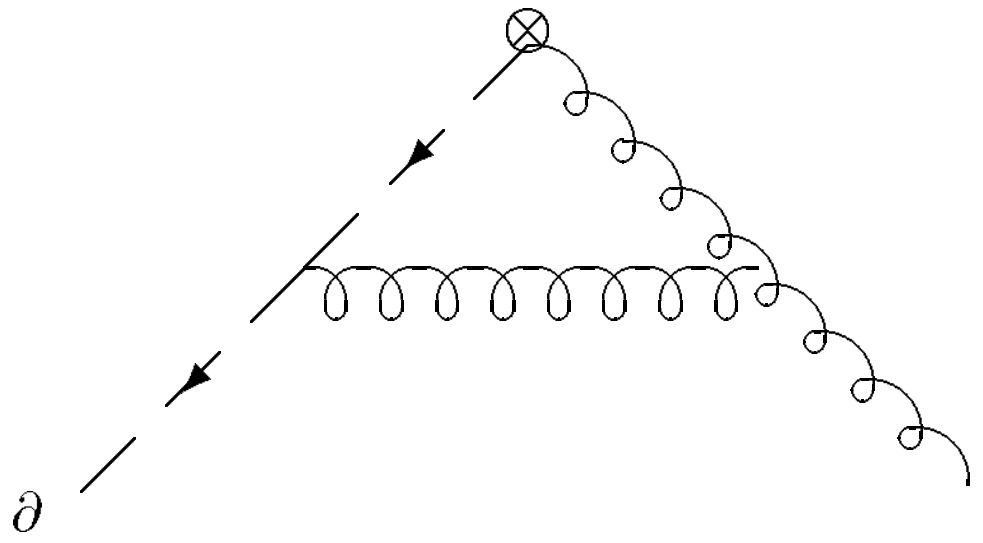

The two graphs above have a derivative acting on the ghost field at the end of the external line.

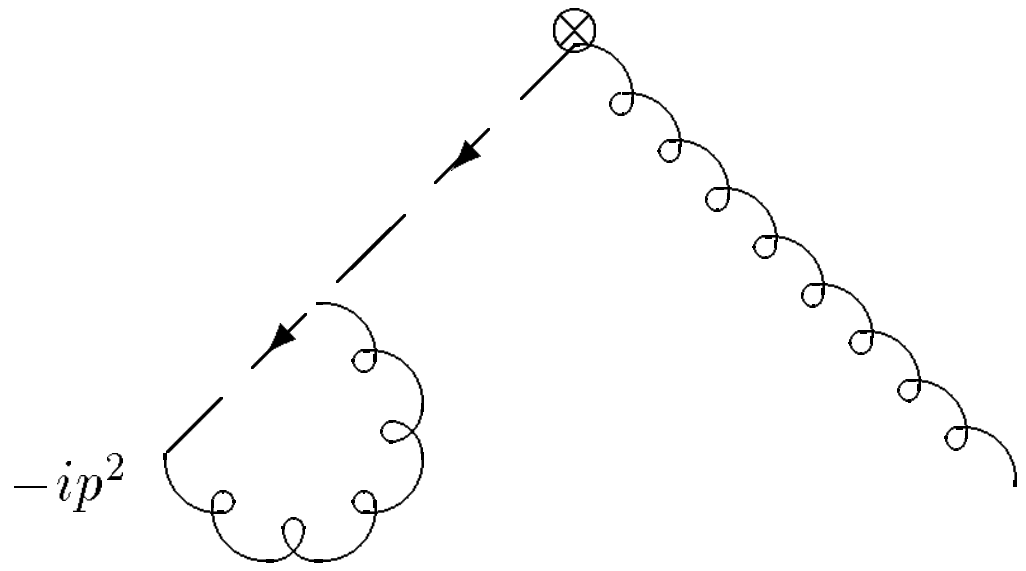

The last two graphs contain an unusual vertex (a vertex not contained in the Lagrangian density), one ghost and one gluon field at the same space-time point. Also, they are each multiplied by the inverse of a ghost propagator as part of the amputation procedure. The three graphs above vanish in physical matrix elements: the two graphs with external derivatives will be proportional to the external momentum and this contracted with the physical gluon polarization will give zero;

the graph immediately above contains a free Lorentz index in its unusual vertex and must be proportional to the external momentum as well. Only the graph below contributes to on-shell matrix elements.

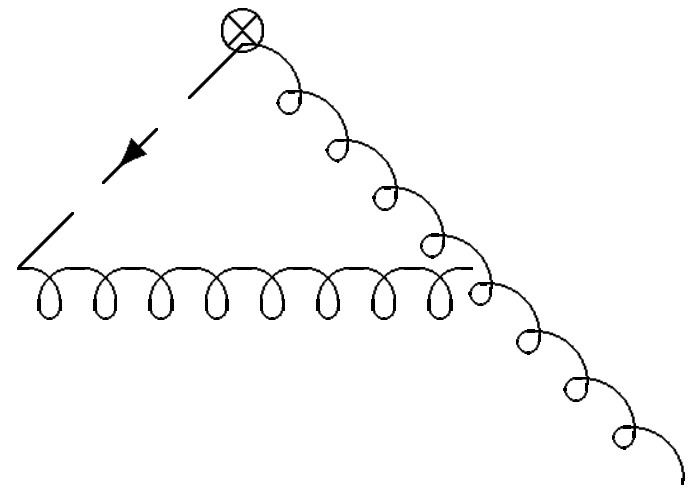

$$
\begin{aligned}
&= \frac{1}{\epsilon}\frac{g^2}{16\pi^2} C_A \delta_{ab} i \left[\frac{1}{12}\left(1-\frac{1}{\lambda}\right)+\frac{1}{4}\right]\left(-\frac{1}{2}g_{\sigma\tau}g_{\mu\nu}+g_{\sigma\mu}g_{\tau\nu}+g_{\sigma\nu}g_{\tau\mu}\right) \\
&+\frac{g^2}{16\pi^2} C_A \delta_{ab} i \left\{
\begin{array}{l}
\frac{1}{4}\left(1-\frac{1}{\lambda}\right)^2 \left[\begin{array}{l}
\frac{p_\sigma p_\tau p_\mu p_\nu}{(p^2)^2}+\frac{1}{2}(-g_{\sigma\tau}g_{\mu\nu}+g_{\sigma\mu}g_{\tau\nu}+g_{\sigma\nu}g_{\tau\mu}) \\
+\frac{p_\mu p_\nu}{p^2}g_{\sigma\tau}-\frac{1}{2}\frac{p_\tau p_\nu}{p^2}g_{\sigma\mu}-\frac{1}{2}\frac{p_\tau p_\mu}{p^2}g_{\sigma\nu} \\
-\left(\frac{p_\sigma p_\nu}{p^2}g_{\tau\mu}+\frac{p_\sigma p_\mu}{p^2}g_{\tau\nu}-\frac{1}{2}\frac{p_\sigma p_\tau}{p^2}g_{\mu\nu}\right)
\end{array}\right] \\
+\left(1-\frac{1}{\lambda}\right)\left[\begin{array}{l}
-\frac{2}{3}\frac{p_\sigma p_\tau p_\mu p_\nu}{(p^2)^2}+\left[\frac{49}{72}+\frac{1}{24}\ln\left(\frac{-p^2}{\bar{\mu}^2}\right)\right]g_{\sigma\tau}g_{\mu\nu} \\
-\left[\frac{23}{72}+\frac{1}{12}\ln\left(\frac{-p^2}{\bar{\mu}^2}\right)\right](g_{\sigma\mu}g_{\tau\nu}+g_{\sigma\nu}g_{\tau\mu}) \\
-\frac{17}{12}\frac{p_\mu p_\nu}{p^2}g_{\sigma\tau}+\frac{1}{3}\frac{p_\tau p_\nu}{p^2}g_{\sigma\mu}+\frac{1}{3}\frac{p_\tau p_\mu}{p^2}g_{\sigma\nu} \\
+\frac{5}{6}\left(\frac{p_\sigma p_\nu}{p^2}g_{\tau\mu}+\frac{p_\sigma p_\mu}{p^2}g_{\tau\nu}-\frac{1}{2}\frac{p_\sigma p_\tau}{p^2}g_{\mu\nu}\right)
\end{array}\right] \\
+\left[-\frac{5}{4}+\frac{1}{8}\ln\left(\frac{-p^2}{\bar{\mu}^2}\right)\right]g_{\sigma\tau}g_{\mu\nu}+\left[\frac{1}{2}-\frac{1}{4}\ln\left(\frac{-p^2}{\bar{\mu}^2}\right)\right](g_{\sigma\mu}g_{\tau\nu}+g_{\sigma\nu}g_{\tau\mu}) \\
+3\frac{p_\mu p_\nu}{p^2}g_{\sigma\tau}-\frac{3}{2}\left(\frac{p_\sigma p_\nu}{p^2}g_{\tau\mu}+\frac{p_\sigma p_\mu}{p^2}g_{\tau\nu}-\frac{1}{2}\frac{p_\sigma p_\tau}{p^2}g_{\mu\nu}\right)
\end{array}\right\} \\
&+\mathcal{O}(g^4)
\end{aligned}
\tag{A.9}
$$

This graph plus its mirror image ($\sigma \leftrightarrow \tau$, $a \leftrightarrow b$, $p \leftrightarrow -p$) contribute to the two amputated

Green functions

$$
\begin{aligned}
&-\left\langle 0\left|T\left[Z_g Z_0 Z_A^{\frac{1}{2}} g c_{ade} A_{\sigma\ d}\omega_e\right] ancestor\left(\theta_{\mu\nu}^{(GV)}\right) A_{\tau\ b}\right|0\right\rangle_{Amputated} \\
&+\left\langle 0\left|T A_{\sigma\ a} ancestor\left(\theta_{\mu\nu}^{(GV)}\right)\left[Z_g Z_0 Z_A^{\frac{1}{2}} g c_{bde} A_{\tau\ d}\omega_e\right]\right|0\right\rangle_{Amputated}
\end{aligned}
\tag{A.10}
$$

corresponding to the unamputated Green functions found in Eq. 4.30. The terms proportional to $\frac{p_\mu p_\nu}{p^2} g_{\sigma\tau}$ in the finite part ruin the proof that the physical matrix element of an alien operator must vanish.[14]

To isolate the parts that survive on shell, we contract Eq. (A.9) with a physical gluon polarization vector, $\epsilon^\tau$, to obtain

$$
\begin{aligned}
&\frac{1}{\epsilon}\frac{g^2}{16\pi^2} C_A \delta_{ab} i\left[\frac{1}{12}\left(1-\frac{1}{\lambda}\right)+\frac{1}{4}\right]\left(-\frac{1}{2} g_{\mu\nu}\epsilon_\sigma + g_{\sigma\mu}\epsilon_\nu + g_{\sigma\nu}\epsilon_\mu\right) \\
&+\frac{g^2}{16\pi^2} C_A \delta_{ab} i \left\{
\begin{array}{l}
\frac{1}{4}\left(1-\frac{1}{\lambda}\right)^2 \left[\begin{array}{l} \frac{1}{2}(-g_{\mu\nu}\epsilon_\sigma + g_{\sigma\mu}\epsilon_\nu + g_{\sigma\nu}\epsilon_\mu) \\ +\dfrac{p_\mu p_\nu}{p^2}\epsilon_\sigma - \left(\dfrac{p_\sigma p_\nu}{p^2}\epsilon_\mu + \dfrac{p_\sigma p_\mu}{p^2}\epsilon_\nu\right) \end{array}\right] \\
+\left(1-\frac{1}{\lambda}\right)\left[\begin{array}{l} \left[\frac{49}{72}+\frac{1}{24}\ln\left(\dfrac{-p^2}{\bar\mu^2}\right)\right] g_{\mu\nu}\epsilon_\sigma \\ -\left[\frac{23}{72}+\frac{1}{12}\ln\left(\dfrac{-p^2}{\bar\mu^2}\right)\right](g_{\sigma\mu}\epsilon_\nu + g_{\sigma\nu}\epsilon_\mu) \\ -\frac{17}{12}\dfrac{p_\mu p_\nu}{p^2}\epsilon_\sigma + \frac{5}{6}\left(\dfrac{p_\sigma p_\nu}{p^2}\epsilon_\mu + \dfrac{p_\sigma p_\mu}{p^2}\epsilon_\nu\right) \end{array}\right] \\
+\left[-\frac{5}{4}+\frac{1}{8}\ln\left(\dfrac{-p^2}{\bar\mu^2}\right)\right] g_{\mu\nu}\epsilon_\sigma + \left[\frac{1}{2}-\frac{1}{4}\ln\left(\dfrac{-p^2}{\bar\mu^2}\right)\right](g_{\sigma\mu}\epsilon_\nu + g_{\sigma\nu}\epsilon_\mu) \\
+3\dfrac{p_\mu p_\nu}{p^2}\epsilon_\sigma - \frac{3}{2}\left(\dfrac{p_\sigma p_\nu}{p^2}\epsilon_\mu + \dfrac{p_\sigma p_\mu}{p^2}\epsilon_\nu\right)
\end{array}
\right\} \\
&+\mathcal{O}(g^4)
\end{aligned}
\tag{A.11}
$$

[14] The pole piece cannot contain a $\frac{1}{p^2}$ divergence because of locality.

## A.5 Feynman Rules

In this section, we give our conventions for the Feynman rules of common objects in pQCD and the non-standard vertices introduced in this paper.

### A.5.1 Propagators

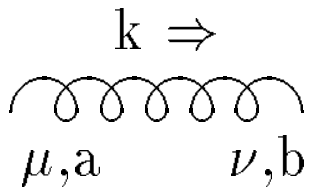

$$= \frac{i\,\delta_{ab}}{k^2 + i\epsilon}\left[-g_{\mu\nu} + \left(1 - \frac{1}{\lambda}\right)\frac{k_\mu k_\nu}{k^2 + i\epsilon}\right] \tag{A.12}$$

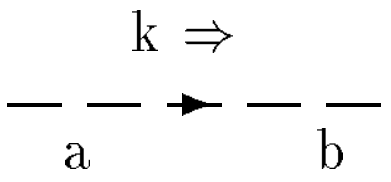

$$= \frac{i\,\delta_{ab}}{k^2 + i\epsilon} \tag{A.13}$$

Include a factor of $-1$ for every ghost loop.

### A.5.2 Vertices in the Lagrangian Density

All momenta are defined to flow *into* the vertex under consideration.

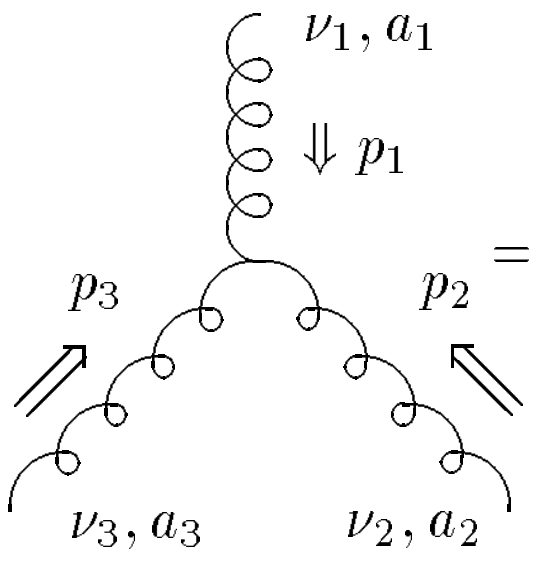

$$= -g c_{a_1 a_2 a_3}[g^{\nu_1\nu_2}(p_1 - p_2)^{\nu_3} + g^{\nu_2\nu_3}(p_2 - p_3)^{\nu_1} + g^{\nu_3\nu_1}(p_3 - p_1)^{\nu_2}] \tag{A.14}$$

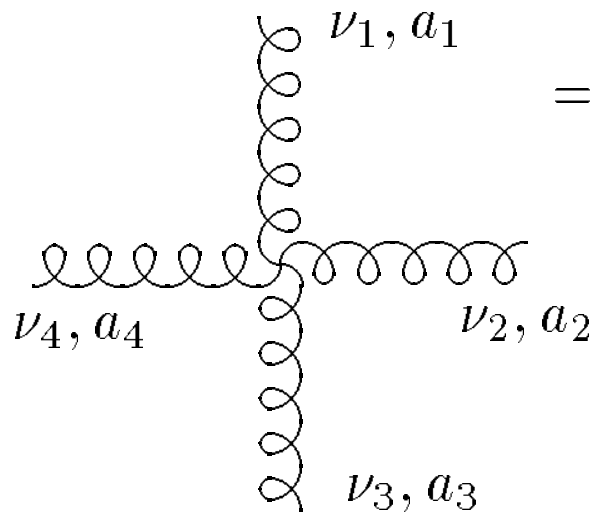

$$\begin{aligned} &= -i\, g^2[c_{ba_1a_2}c_{ba_3a_4}(g^{\nu_1\nu_3}g^{\nu_2\nu_4} - g^{\nu_1\nu_4}g^{\nu_2\nu_3}) \\ &\quad + c_{ba_1a_3}c_{ba_4a_2}(g^{\nu_1\nu_4}g^{\nu_3\nu_2} - g^{\nu_1\nu_2}g^{\nu_3\nu_4}) \\ &\quad + c_{ba_1a_4}c_{ba_2a_3}(g^{\nu_1\nu_2}g^{\nu_4\nu_3} - g^{\nu_1\nu_3}g^{\nu_4\nu_2})] \end{aligned} \tag{A.15}$$

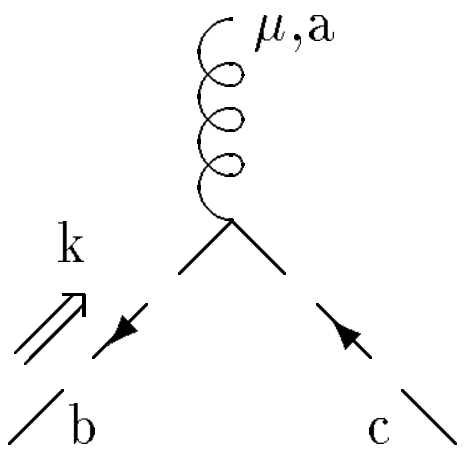

$$= g c_{abc} k^{\mu} \tag{A.16}$$

### A.5.3 Non-Standard Vertices at Zero Momentum Transfer

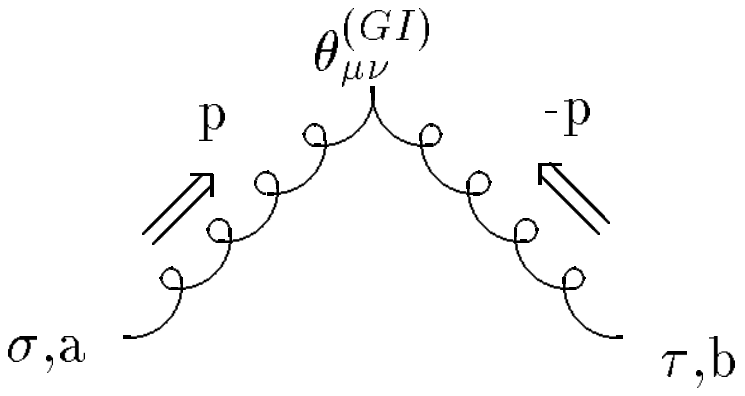

$$= \begin{array}{l} \delta ab[p^2(g_{\sigma\tau}g_{\mu\nu} - g_{\sigma\mu}g_{\tau\nu} - g_{\sigma\nu}g_{\tau\mu}) - 2p_\mu p_\nu g_{\sigma\tau} \\ +p_\tau p_\nu g_{\sigma\mu} + p_\tau p_\mu g_{\sigma\nu} + p_\sigma p_\nu g_{\tau\mu} + p_\sigma p_\mu g_{\tau\nu} - p_\sigma p_\tau g_{\mu\nu}] \end{array} \tag{A.17}$$

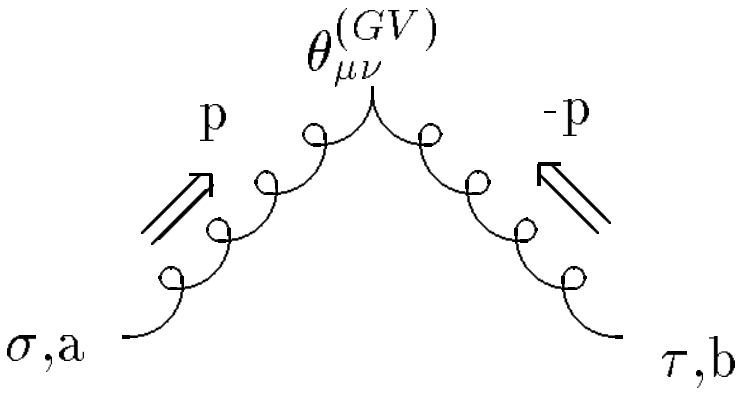

$$= -\delta_{ab}\lambda(p_\tau p_\nu g_{\sigma\mu} + p_\tau p_\mu g_{\sigma\nu} + p_\sigma p_\nu g_{\tau\mu} + p_\sigma p_\mu g_{\tau\nu} - p_\sigma p_\tau g_{\mu\nu}) \tag{A.18}$$

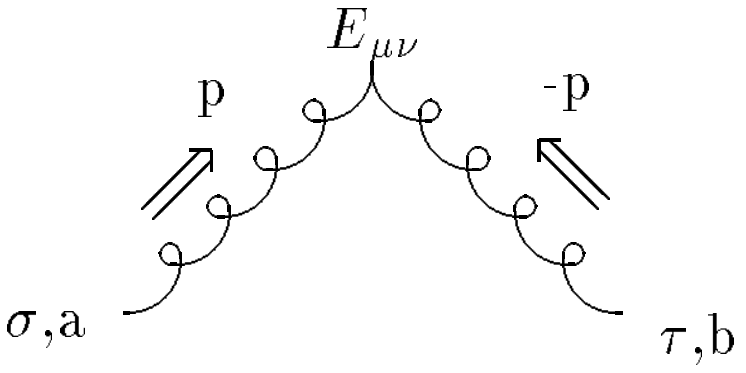

$$= \begin{array}{l} \delta_{ab}[-2p^2(-\frac{1}{2}g_{\sigma\tau}g_{\mu\nu}g_{\sigma\mu}g_{\tau\nu} + g_{\sigma\nu}g_{\tau\mu}) \\ +(1-\lambda)(p_\tau p_\nu g_{\sigma\mu} + p_\tau p_\mu g_{\sigma\nu} + p_\sigma p_\nu g_{\tau\mu} + p_\sigma p_\mu g_{\tau\nu} - p_\sigma p_\tau g_{\mu\nu})] \end{array} \tag{A.19}$$

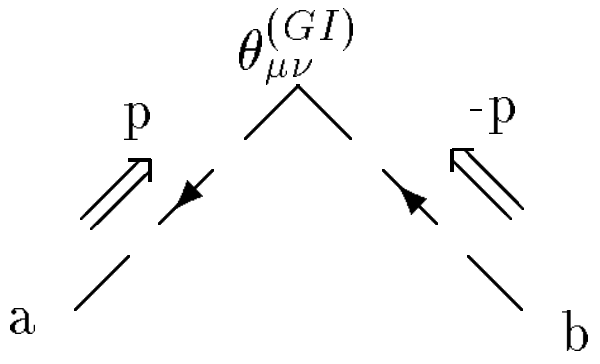

$$= 0 \tag{A.20}$$

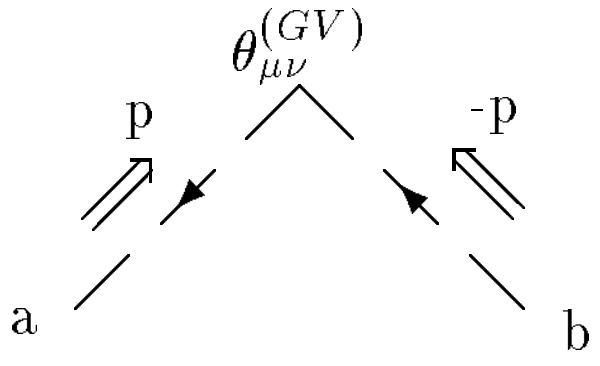

$$= \delta_{ab}(2p_\mu p_\nu - p^2 g_{\mu\nu}) \tag{A.21}$$

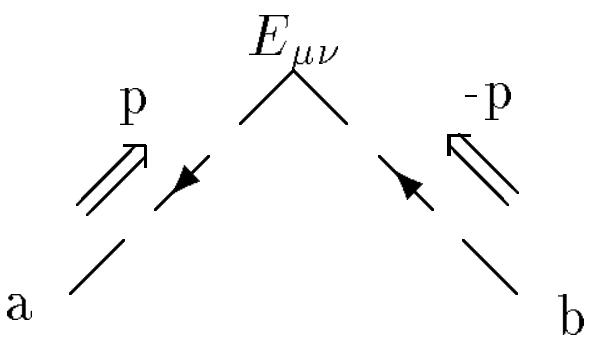

$$= -\tfrac{1}{2}\delta_{ab}p^2 g_{\mu\nu} \qquad \text{(A.22)}$$

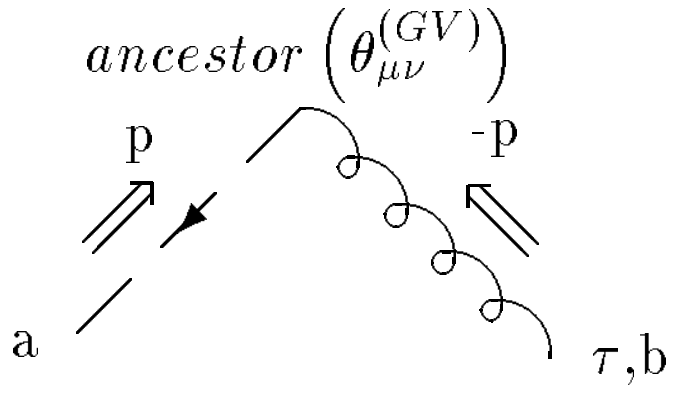

$$= -i\delta_{ab}(p_\nu g_{\mu\tau} + p_\mu g_{\nu\tau} - \tfrac{1}{2}g_{\mu\nu}p_\tau) \qquad \text{(A.23)}$$

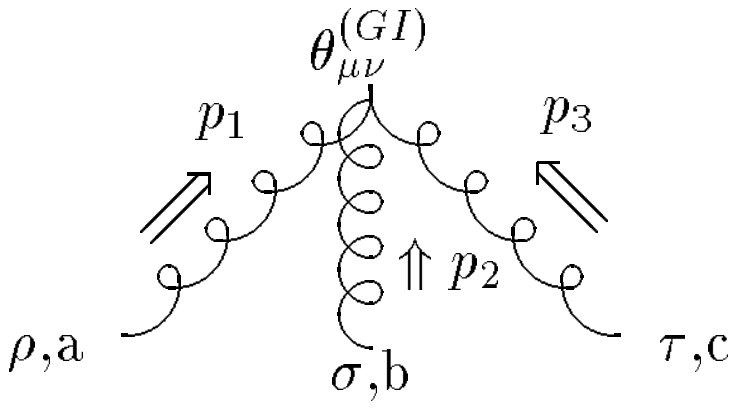

$$= -igc_{abc}\left\{\begin{array}{c} p_{1\mu}g_{\rho\tau}g_{\sigma\nu} - p_{1\tau}g_{\rho\mu}g_{\sigma\nu} + p_{1\nu}g_{\rho\tau}g_{\sigma\mu} - p_{1\tau}g_{\rho\nu}g_{\sigma\mu} \\ -p_{1\mu}g_{\rho\sigma}g_{\tau\nu} + p_{1\sigma}g_{\rho\mu}g_{\tau\nu} - p_{1\nu}g_{\rho\sigma}g_{\tau\mu} + p_{1\sigma}g_{\rho\nu}g_{\tau\mu} \\ +p_{2\mu}g_{\sigma\rho}g_{\tau\nu} - p_{2\rho}g_{\sigma\mu}g_{\tau\nu} + p_{2\nu}g_{\sigma\rho}g_{\tau\mu} - p_{2\rho}g_{\sigma\nu}g_{\tau\mu} \\ -p_{2\mu}g_{\sigma\tau}g_{\rho\nu} + p_{2\tau}g_{\sigma\mu}g_{\rho\nu} - p_{2\nu}g_{\sigma\tau}g_{\rho\mu} + p_{2\tau}g_{\sigma\nu}g_{\rho\mu} \\ +p_{3\mu}g_{\tau\sigma}g_{\rho\nu} - p_{3\sigma}g_{\tau\mu}g_{\rho\nu} + p_{3\nu}g_{\tau\sigma}g_{\rho\mu} - p_{3\sigma}g_{\tau\nu}g_{\rho\mu} \\ -p_{3\mu}g_{\tau\rho}g_{\sigma\nu} + p_{3\rho}g_{\tau\mu}g_{\sigma\nu} - p_{3\nu}g_{\tau\rho}g_{\sigma\mu} + p_{3\rho}g_{\tau\nu}g_{\sigma\mu} \\ +g_{\mu\nu}[g_{\rho\sigma}(p_{1\tau} - p_{2\tau}) + g_{\sigma\tau}(p_{2\rho} - p_{3\rho}) + g_{\tau\rho}(p_{3\sigma} - p_{1\sigma})] \end{array}\right\} \qquad \text{(A.24)}$$

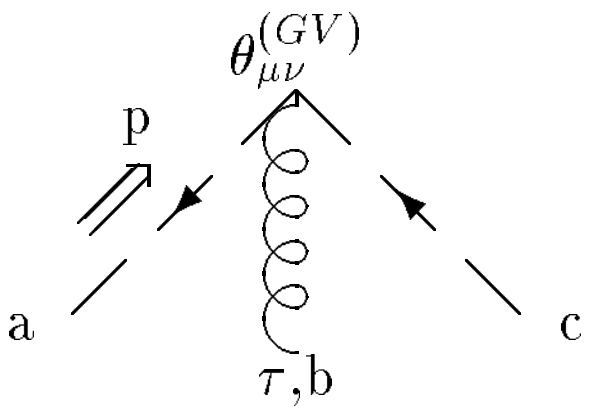

$$= igc_{abc}(p_\mu g_{\nu\tau} + p_\nu g_{\mu\tau} - p_\tau g_{\mu\nu}) \qquad \text{(A.25)}$$